\documentclass[fleqn,usenatbib]{mnras}

\usepackage{newtxtext,newtxmath}
\usepackage[T1]{fontenc}

\DeclareRobustCommand{\VAN}[3]{#2}
\let\VANthebibliography\thebibliography
\def\thebibliography{\DeclareRobustCommand{\VAN}[3]{##3}\VANthebibliography}

\usepackage{graphicx}
\usepackage{amsmath}
\usepackage{siunitx}
\usepackage{dirtytalk}
\usepackage{enumitem}
\usepackage{nth}
\usepackage{CJKutf8}
\usepackage{orcidlink}
\usepackage{placeins}
\usepackage[normalem]{ulem}

\newcommand{\object}[1]{#1}

\title[SN\,2024cld]{SN\,2024cld: unveiling the complex mass-loss histories of evolved supergiant progenitors to core collapse supernovae}

\author[Killestein et al.]{
T.~ L. Killestein$^{1,2}$\thanks{tom.killestein@gmail.com}\orcidlink{0000-0002-0440-9597},
M. Pursiainen$^{1}$\orcidlink{0000-0003-4663-4300},
R. Kotak$^{2}$\orcidlink{0000-0001-5455-3653},
P. Charalampopoulos$^{2,3}$\orcidlink{0000-0002-0326-6715},
J. Lyman$^{1}$,
K. Ackley$^{1}$,
\newauthor
S. Belkin$^{4}$,
D. L. Coppejans$^{1}$\orcidlink{0000-0001-5126-6237},
B. Davies$^{1}$,
M.~J. Dyer$^{5,6}$\orcidlink{0000-0003-3665-5482},
L. Galbany$^{7,8}$,
B. Godson$^{1}$,
D. Jarvis$^{5}$,
\newauthor
N. Koivisto$^{2}$\orcidlink{0009-0007-7151-7313},
A. Kumar$^{9}$\orcidlink{0000-0002-4870-9436},
M. Magee$^{1}$,
M. Mitchell$^{1}$,
D. O'Neill$^{10,11}$,
A. Sahu$^{1}$,
B. Warwick$^{1}$,
\newauthor
R.~P. Breton$^{12}$,
T. Butterley$^{13}$,
Y.-Z. Cai~\begin{CJK*}{UTF8}{gbsn}(蔡永志)\end{CJK*}$^{14,15}$\orcidlink{0000-0002-7714-493X},
J. Casares$^{16}$,
V.~S. Dhillon$^{5,16}$,
N. Elias-Rosa$^{17,7}$,
\newauthor
M. Fraser$^{18}$,
D.~K. Galloway$^{4}$,
B. Gompertz$^{10,11}$,
M. González-Bañuelos$^{7,8}$\orcidlink{0009-0006-6238-3598},
C.~P. Gutiérrez$^{8,7}$\orcidlink{0000-0003-2375-2064},
\newauthor
T. Kangas$^{3,2}$,
E. Kankare$^{2}\orcidlink{0000-0001-8257-3512}$,
L. Kelsey$^{19}$,
T. Kravtsov$^{2}$,
G. Leloudas$^{20}$,
S.~P. Littlefair$^{5}$,
K. Matilainen$^{2}$,
\newauthor
S. Mattila$^{2,21}$,
T. Nagao$^{2,22,23}$,
K. Noysena$^{24}$,
L.~K. Nuttall$^{25}$,
P. O'Brien$^{26}$,
D. Pollacco$^{1}$,
G. Ramsay$^{27}$,
\newauthor
A. Reguitti$^{28,17}$,
T.~M. Reynolds$^{29,30,31}$\orcidlink{0000-0002-1022-6463},
I. Salmaso$^{32,17}$\orcidlink{0000-0003-1450-0869},
R.~L.~C. Starling$^{26}$,
D. Steeghs$^{1}$,
M. Stritzinger$^{33}$,
\newauthor
K. Ulaczyk$^{1}$,
G. Valerin$^{17}$,
Z.-Y. Wang$^{34,35}$\orcidlink{0000-0002-0025-0179},
R. Wilson$^{13}$
\\
(Affiliations can be found after the references)
}

\date{Accepted 2025 December 19. Received 2025 December 15; in original form 2025 October 31}

\pubyear{\the\year{}}

\begin{document}
\label{firstpage}
\pagerange{\pageref{firstpage}--\pageref{lastpage}}
\maketitle

\begin{abstract}
Pre-explosion mass loss in supernova (SN) progenitors is a crucial unknown factor in stellar evolution, yet has been illuminated recently by the diverse zoo of interacting transients. We present \object{SN\,2024cld}, a transitional core-collapse SN at a distance of 39\,Mpc, straddling the boundary between SN~II and SN~IIn, showing persistent interaction with circumstellar material (CSM) similar to H-rich \object{SN\,1998S} and \object{PTF11iqb}. The SN was discovered and classified just 12\,h post-explosion via the GOTO-FAST high-cadence program. Optical spectroscopy, photometry, and polarimetry over 220\,d chart the complex, long-lived interaction in this transient. Early evolution is dominated by CSM interaction, showing a 14\,d rise to a peak absolute magnitude of $g=-17.6$ mag, with clear flash-ionisation signatures. SN\,2024cld also shows a slowly-evolving late time light curve powered by CSM interaction, with high-velocity (\qty{6000}{\kilo\meter\per\second}) shoulders on a strong multi-component H$\alpha$ profile. Dense polarimetric coverage reveals marked evolution in the photospheric geometry -- peaking at 2\% polarisation 10 days post-explosion, and rotating \qty{\approx60}{\degree} as the ejecta sweep more distant CSM. We observe  a narrow (\qty{\approx 60}{\kilo\meter\per\second}) H$\alpha$ P Cygni feature throughout, associated with pre-shock CSM. SN\,2024cld represents among the best-observed 98S-like SNe to date, revealing a multi-component CSM structure: a dense, inner aspherical envelope, CSM disk/torus, and tenuous, extended wind. We propose this SN arose from an evolved supergiant progenitor experiencing multiple mass loss episodes in its terminal years, with binary interaction plausibly generating the CSM disk. SN\,2024cld constrains the progenitors and mass-loss paradigms of 98S-like SNe, unveiling the chaotic ends of evolved supergiant stars from afar.
\end{abstract}

\begin{keywords}
    supernovae: general --
    stars: circumstellar matter --
    supernovae: individual: SN\,2024cld
\end{keywords}

\section{Introduction}
    Mass loss in the final years of the lives of massive stars remains a poorly-understood, yet crucial phenomenon for driving the latter phases of stellar evolution, seeding the interstellar medium \citep[ISM; ][]{Leitherer1992}, and generating the diverse set of transients we observe~\citep[e.g.][]{Woosley2002,Smith2014}. Transients exploding within dense circumstellar material (CSM), known as interacting transients (\citealt{Schlegel1990}, see \citealt[][]{Fraser2020} for a recent review), represent the most direct probes of pre-explosion mass loss, as the expanding ejecta collides with recently-ejected material, providing an additional energy source via shock heating, thus driving the light curve and spectral evolution of these events in complex ways.
    
    Observational evidence highlights the significance of mass loss across hydrogen-rich supernovae (SNe), including in the most common Type II SNe (SNe~II), a class that arise from red supergiant (RSG) progenitors. The early-time light-curves of SNe~II are found to rise more quickly~\citep{Gall2015,Gonzalez-Gaitan2015} than predicted by hydrodynamical models of their progenitors, not trivially explained by fine-tuning progenitor radii, constrained from direct detections of SN\,II progenitors~\citep[e.g. ][]{Smartt2009,O'Neill2019}. The faster rise times are best explained via either a tenuous highly-extended envelope, or dense, confined CSM close-in~\citep[e.g. ][]{Morozova2018} to the progenitor. Early-time spectroscopy provides further evidence for the presence of this CSM through narrow emission lines. These arise from the recombination of the CSM, ionised either by emission from the shock breakout/cooling of the SN\,\citep[flash ionisation, e.g. ][]{Gal-Yam2014,Khazov2016,Yaron2017}, or from the shocks at the ejecta-CSM boundary. This necessitates CSM that is both close-in to the progenitor, and of sufficient density to generate the strong emission seen. Early-time spectroscopy~\citep{Khazov2016} is a powerful diagnostic of the CSM properties, showing considerable diversity in composition and ionisation state~\citep[e.g. ][]{Gal-Yam2014,Tartaglia2021,Terreran2022,Bostroem2023,Jacobson-Galan2024} and their evolution. Recent magnitude-limited samples~\citep{Bruch2023} suggest $\gtrsim30\%$ of SNe~II show flash ionisation (FI) signatures within the first few days of explosion -- indicating that enhanced late-time mass loss must be common among SN progenitors. 

    In spite of this rich picture of the terminal mass loss of core collapse SNe (CCSNe) from infant transient discovery,  theoretical gaps remain in our understanding of late-stage stellar evolution, largely around the mass loss mechanisms themselves. Flash spectroscopy of larger samples of CCSNe points (broadly) towards mass loss rates of $10^{-3}-10^{-2}\, \mathrm{\mathrm{M_\odot}\,yr^{-1}}$ assuming typical RSG wind velocities, commencing $\sim10-100$\,yr prior to explosion~\citep{Jacobson-Galan2024}. 
    It has been suggested that enhanced winds prior to the SN explosion dubbed \say{superwinds}~\citep[e.g. ][]{Moriya2017}, driven by turbulence~\citep{Kee2021} or pulsations in the progenitor~\citep{Yoon2010}
    may account for this timescale. It is notable that this inferred mass loss rate is orders of magnitude higher than seen in even the most extreme Galactic RSGs~\citep{Beasor2018} -- suggesting the mass loss must occur on much shorter timescales, and be shed far more quickly than assumed. This mechanism appears at odds with the observed properties of SN\,II progenitors, Galactic RSGs, which do not show as significant dust obscuration~\citep{Davies2022} as should be expected from the more prolonged mass loss inferred in SNe. Recent radio studies~\citep{Sfaradi2025} also strongly disfavour these inferred mass loss rates in a wind-driven context. Eruptive mass loss events in massive stars are thought to be behind the prodigious levels of CSM interaction seen in strongly-interacting SNe~II~\citep[known as SNe~IIn, e.g.][]{Dwarkadas2011,Smith2017_LBVs} and related types, evidenced by the strong \say{precursor} emission in the years prior to explosion observed in a number of such objects further cementing this~\citep[e.g.][]{Ofek2014, Strotjohann2021}. It remains unclear what fraction of SN~II progenitors experience these eruptions prior to explosion -- some plausible precursor outbursts have been observed~\citep[e.g. ][]{Ho2019,Jacobson-Galan2022,Reguitti2024}, but the sample remains modest owing to the relative faintness of the precursors.

    Further complicating the progenitor's terminal evolution is the fact that the majority of massive stars exist within binary or multiple star systems \citep[e.g.][]{Sana2012}, enabling envelope-stripping of the progenitor to further modulate the surrounding CSM in complex ways~\citep{Podsiadlowski1992} -- through binary mass shedding events or common-envelope evolution. The presence of a companion can strongly shape the geometry of the ejected mass, creating asymmetric CSM configurations that reveal themselves in (spectro-)polarimetry of the SN~\citep{Wang2001} and skewed line profiles~\citep[e.g. ][]{Taddia2020}. Binary mass transfer~\citep[e.g.][]{Ercolino2025} may also occur episodically, naturally yielding a stratified and complex CSM configuration that is difficult to explain via single-star evolution -- at least not without invoking multiple mechanisms simultaneously. This is highlighted by transitional CCSNe that straddle the line between IIn and slow- and fast-declining SNe~II (SNe~IIP and SNe~IIL respectively) through their interaction signatures, and whose peculiar inferred properties (prolonged interaction, disk-like CSM, discrete CSM shells, marked asymmetry) are not straightforwardly explained by simpler mass loss routes. \object{SN\,1998S}~\citep[e.g.][]{Fassia2000,Fassia2001} and \object{PTF11iqb}~\citep{Smith2015} are two such SNe in this class. Both objects initially showed narrow emission lines of high-ionisation species, with broad wings, which is indicative of flash-ionised CSM. After a considerable delay, they showed the broad H$\alpha$ emission typical of IIP/IIL SNe -- albeit without P Cygni profiles, which were presumably filled in by interaction~\citep{Dessart2016}. The light curve of PTF~11iqb was broadly similar to that of SNe~IIP, showing a plateau from days 50-120~\citep{Smith2015}, whereas SN\,1998S showed a more IIL-like decline between days 30-70~\citep{Fassia2000}, and was markedly brighter. The two objects were spectrally very similar, in spite of this divergent light-curve behaviour. Towards later times, both objects showed significant evolution in the H$\alpha$ profile, developing multiple distinct components at high ($\sim1000$s km/s) velocity, and still retaining narrow H$\alpha$ emission in the transient rest frame that was observed since explosion. These multi-component, complex H emission profiles point towards a complex, density-stratified CSM, likely with marked asymmetry considering the elevated ($\approx2\,\%$) polarisation seen in both SN\,1998S~\citep{Leonard2000} and PTF11iqb~\citep{Bilinski2024}. Given the distinct CSM structure, prior works~\citep{Smith2015} have appealed to multiple shedding events induced by a binary companion~\citep{Scherbak2025,Ercolino2025}, as hinted above, to generate a toroidal/disk-like CSM distribution around the progenitor.
        
    Whether the observed properties of SNe~IIL, IIP, IIn, and everything in between can be explained solely by increasing the amount of mass loss~\citep[e.g. ][]{Smith2014}, or whether distinct and unique mass loss mechanisms and progenitors are responsible for each remains an open question. Given that these mass-loss mechanisms produce very similar \citep[often degenerate in mass loss parameter space, see][]{Dessart2023} observable features in regular SNe~II, it is crucial to obtain datasets that span the early-to-late time evolution with higher-resolution spectroscopy and polarimetry, to better distinguish CSM features/geometries and enable stronger constraints. Observing SNe~II and SNe~IIn close to the boundaries between these traditional classes is crucial for providing observational constraints on the mass loss parameter space, and thus establishing the dominant mechanisms at play from the CSM configuration observed.
    In this paper we present an in-depth spectrophotometric and polarimetric study of the transitional SN~II SN\,2024cld, spanning the first 200 days of evolution, and unveiling the complex mass-loss history of the progenitor star through enduring CSM interaction.

    This paper is organised as follows: In Section~\ref{sec:observations} we outline the extensive dataset collated on SN\,2024cld. Section~\ref{sec:hostglx} details the host galaxy and local environment of SN\,2024cld. In Sections~\ref{sec:photometry} and \ref{sec:spectroscopy} we present a comprehensive analysis of the spectroscopic and photometric properties. Section~\ref{sec:polarimetry} discusses the long-term polarimetric series obtained. Section~\ref{sec:discussion} ties together all observational data, discusses the unique characteristics of SN\,2024cld, and conjectures the potential progenitor systems compatible with the observed properties.

\section{Observations and data reduction} \label{sec:observations}

\begin{figure}
        \centering
        \includegraphics[width=\linewidth]{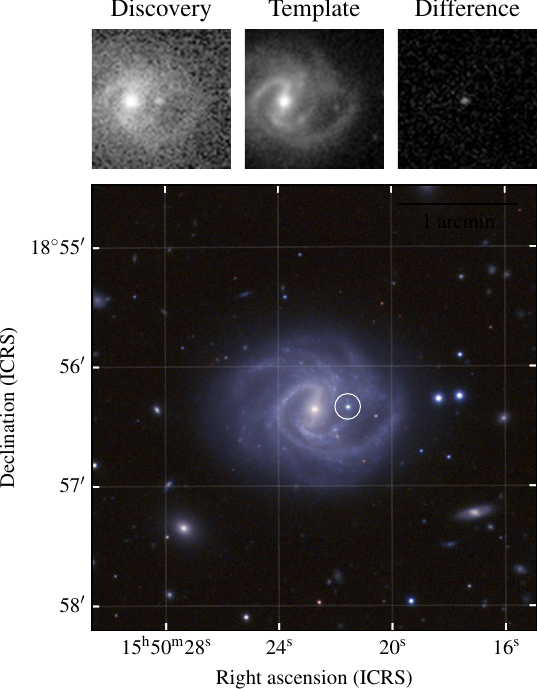}
        \caption{NOT/ALFOSC $gri$ deep stack showing SN\,2024cld (circled), embedded in the spiral arm of host galaxy NGC 6004. The panels above show the GOTO discovery/template/difference image for SN\,2024cld.}
        \label{fig:2024cld_leadpic}
    \end{figure}

    SN\,2024cld, located at $\alpha=\mathrm{15^h50^m21\overset{s}{.}55, \delta=+18^{\circ}56'20\overset{\arcsec}{.}22}$ \citep[from Gaia Alerts; ][]{Hodgkin2021}, was discovered by the Gravitational-wave Optical Transient Observer \citep[GOTO; ][]{Steeghs2022, Dyer2024} as part of the dedicated GOTO Fast Analysis and Spectroscopy of Transients~\citep[GOTO-FAST;][]{Godson2023Intro} high-cadence survey. SN\,2024cld, assigned the internal name GOTO\,24ql, was discovered on 2024 February 13 05:28:41 UT at a magnitude of $L=17.97$. The discovery triplet of science, template, and difference images from GOTO, and a deep multi-colour image of SN\,2024cld are shown in Figure~\ref{fig:2024cld_leadpic}. 
    The closest non-detection was 2024 February 12 06:25:27 UT, to a depth of $L>19.1$ mag, constraining the explosion epoch to within 23.3\,h of discovery. A prompt spectrum~\citep{cld_classification} of SN\,2024cld was initiated just 28 mins post-discovery with the \qty{2.54}{\meter} Isaac Newton Telescope (INT) on La Palma, and revealed a blue continuum with strong, narrow lines of H, and a broad blend of He~II, N~III, and C~III. Based on this, \citet{Warwick2024} classified SN\,2024cld as a flash-ionised SN\,II at $z\approx0.014$,\footnote{The redshift we assume in this work ($z=0.01252$) is lower than reported in the initial classification report owing to a re-reduction of the first spectrum and further medium-resolution spectroscopy (see Section~\ref{sec:hostglx}).}. Continued follow-up revealed an evolution very different to that expected of a typical CCSN.
    
    Photometry and spectroscopy in this paper have been corrected for a galactic reddening of $E_{(B-V)} = 0.1076$, from the dust maps of \cite{Schlafly2011}, and the host reddening (discussed in Section~\ref{sec:hostglx}). We adopt a redshift of $z=0.01252$, based on fits to the H$\alpha$ and [N II] emission from the underlying stellar population at the SN explosion site (see Section~\ref{sec:hostglx}). All magnitudes are given in the AB system~\citep{OkeGunnAB}. Times and phases, unless stated, are given in rest-frame days relative to the inferred explosion epoch. Throughout the paper, we also assume the \citet{PlanckCollaboration2020} $\Lambda$CDM cosmology ($H_0=67.66$\,km\,s$^{-1}$\,Mpc$^{-1}$, $\Omega_m=0.310$).

    \subsection{Survey photometry}
        We collated and curated survey photometry from the GOTO, Asteroid Terrestrial-impact Last Alert System~\citep[ATLAS; ][]{Tonry2018}, Zwicky Transient Facility~\citep[ZTF; ][]{Bellm2019}, and All-Sky Automated Survey for Supernovae~\citep[ASAS-SN; ][]{Shappee2014} surveys, via their respective forced photometry services: the GOTO forced photometry service (Jarvis et al., in prep.), the ATLAS Forced Photometry Server~\citep{Shingles2021}, the ZTF Forced Photometry Service~\citep{Masci2023}, and the ASAS-SN Sky Patrol~\citep{Kochanek2017} website. The same post-processing routines were applied to the outputs of each service: rejection of poorly-subtracted/deviant epochs via $\sigma$-clipping, baseline correction using pre-explosion photometry, and stacking of multiple visits on the same night into single flux measurements to maximise signal-to-noise. We applied an additional colour correction to the GOTO photometry given the significant colour evolution of SN\,2024cld, removing colour terms and calibrating it to the ATLAS REFCAT2 $g$-band~\citep{Refcat2}.

    \subsection{Optical photometry} \label{sec:opticalphot}
        SN\,2024cld was observed with a number of instruments: IO:O on the \qty{2}{\meter} Liverpool Telescope (LT; \citealt{Steele2004}) in $ugriz$, Rapid-Eye Mount telescope (REM; \citealt{Chincarini2003}) in $griz$, Alhambra Faint Object Spectrograph and Camera (ALFOSC) mounted on the \qty{2.56}{\meter} Nordic Optical Telescope (NOT) in $ugriz$ via the NOT Un-biased Transient Survey (NUTS2\footnote{\url{https://nuts2.sn.ie}}), pt5m~\citep{Hardy2015} in $BVRI$, Asiago Faint Objects Spectrograph and Camera (AFOSC) on the Copernico 1.92m telescope in $ugriz$, and Asiago Schmidt 67/92 telescope in $uBVgri$. All imaging data were reduced with a custom pipeline, primarily using seeing-matched aperture photometry on difference images to mitigate issues with poor point spread function (PSF) reconstruction and sampling. Given the bright host galaxy, the difference imaging process was carefully optimised to yield the cleanest subtractions possible, using the \texttt{HOTPANTS}~\citep{Becker2015} algorithm -- with particular care paid to the background estimation and propagation of variance. We used masks based on the Siena Galaxy Atlas~\citep{Moustakas2023} morphological measurements to avoid over-subtraction of galaxy light in sky background subtraction, and to minimise the influence of contaminated stars on the zero-point estimation. For the host galaxy templates, we used Sloan Digital Sky Survey Data Release 16 (SDSS DR16;~\citealt{Ahumada2020}) $ugriz$-band stacks. We derived synthetic $BVRI$ magnitudes for sequence stars, and synthetic $BVRI$ templates for subtractions in these filters using the Lupton (2005) transformation equations to create correctly-weighted combinations of SDSS $gri$ template images. All magnitudes given in this paper are on the AB system~\citep{OkeGunnAB}. All photometry, including in sections that follow, is summarised in Table~\ref{tab:photlog}.

    \subsection{Infrared photometry}
        Near-infrared imaging was performed with the Nordic Optical Telescope near-infrared Camera and spectrograph (NOTCam) via NUTS2, in the $JHK$ bands, and reduced using custom IRAF-based routines. $H$-band imaging was also obtained with REM, and reduced with the standard REM pipeline. Photometry was derived following the same reduction steps, and using the same custom photometric pipeline as in Section~\ref{sec:opticalphot},  using Two Micron All Sky Survey \citep[2MASS; ][]{Skrutskie2006} stars as local calibrators to derive the zero points. Images from the United Kingdom Infra-Red Telescope (UKIRT) Hemisphere Survey (UHS) DR3~\citep{Dye2018} were used as templates for difference imaging.
        
        \subsection{UV photometry and X-ray upper limits}
        The Ultraviolet-Optical Telescope (UVOT; \citealt{Roming2005}) aboard the Neil Gehrels \textit{Swift} Observatory observed SN\,2024cld in \textit{UVW2, UVM2, UVM1, U, B,} and \textit{V} bands as a target of opportunity (PIs: Killestein, Jacobson-Galán), receiving 14 visits in total.
        The \textit{Swift} UVOT data are reduced following the standard procedures in \citet{Charalampopoulos2024}. Commensal \textit{Swift} X-ray Telescope (XRT) observations were also performed during each visit, in Photon Counting (PC) mode.  No concrete X-ray detection of SN\,2024cld was made, with typical upper limits of $10^{-2}$ counts/s in the 0.3-10 keV band. We estimated $2\sigma$ upper limits on the X-ray flux in the \textit{Swift} bandpass using the HILIGT utility~\citep{Saxton2022}, assuming a power-law spectrum with spectral index 2, and a neutral hydrogen column density of $3\times10^{20}$ cm$^{-2}$. We elaborate on these non-detections, and place them into the wider context of high-energy emission from CCSNe in Section~\ref{sec:discussion}.

    \subsection{Optical spectroscopy}
        A dense series of optical spectroscopy was obtained with a number of facilities, covering the first \qty{220}{\day} of the lifetime of SN\,2024cld, which are discussed in chronological order below. Spectra were obtained with the INT Intermediate Dispersion Spectrograph (IDS) using the R150V, R300V, and R400V gratings as part of GOTO-FAST. INT data were reduced with PypeIt~\citep{Prochaska2020}, making use of custom recipes.
        We obtained a number of spectra with the SPectrograph for the Rapid Acquisition of Transients (SPRAT; \citealt{Piascik2014}) on the LT. SPRAT data were reduced with PypeIt as with INT/IDS data. Data were obtained with NOT/ALFOSC (via NUTS2) using grisms 4, 8, 17, and 18 to obtain both medium and low-resolution spectroscopy. The gr4 data were reduced using a custom IRAF-based pipeline (FOSCGUI\footnote{\url{https://sngroup.oapd.inaf.it/foscgui.html}}). 
        All remaining data were reduced using 
        PypeIt and the PyNOT\footnote{\url{https://jkrogager.github.io/pynot/}} package.
        A number of higher-resolution spectra were obtained with the Wide Field Spectrograph (WiFeS; \citealt{Dopita2007}) mounted on the ANU2.3m telescope, using both B/R3000 and B/R7000 gratings. ANU2.3m/WiFeS data were reduced with the pyWiFeS pipeline~\citep{Childress2014}. One spectrum was obtained with OSIRIS+~\citep{Cepa2000} on the \qty{10.4}{\meter} Gran Telescopio Canarias (GTC), using the B/R1000 gratings. All spectra were flux-calibrated using standard star observations, and rescaled to contemporaneous photometry to correct for slit losses. Where standards were taken at sufficiently similar airmass we also corrected for telluric absorption.
        A full log of observations, alongside ancillary data, are given in Table~\ref{tab:spectralog}. All spectra are corrected for Galactic and host reddening assuming Milky Way-like dust ($R_V=3.1$), in absence of any constraints from our dataset on this.
    
    \subsection{Imaging polarimetry}
        A sequence of linear polarimetry for SN\,2024cld was obtained starting at \qty{+3.5}{\day} post-explosion and extending to \qty{+120}{\day}, with NOT/ALFOSC using a combination of Bessell B and Johnson V filters and the $\lambda/2$ waveplate. Polarimetric data is processed following \citet{Pursiainen2023_HRich}, and corrected for Milky Way interstellar polarisation (ISP) using field stars.

\section{Host galaxy and environment} \label{sec:hostglx}
NGC~6004, the host galaxy of SN\,2024cld, is a face-on barred spiral at a distance of 39.3\,Mpc ($m-M = 32.97\pm0.64$ mag), estimated via the Tully-Fisher relation~\citep{Springob2009}. SN\,2024cld exploded on one of the galaxy's spiral arms, projected 17\arcsec (5.3\,kpc) from the nucleus. NGC~6004 also hosted one other observed SN, PS16bkl~\citep{Halevi2016}, classified as a SN\,IIP. PS16bkl is located 55\arcsec (17.5\,kpc) away from SN\,2024cld, also on a spiral arm of the host.

\begin{figure}
    \centering
    \includegraphics[width=\linewidth]{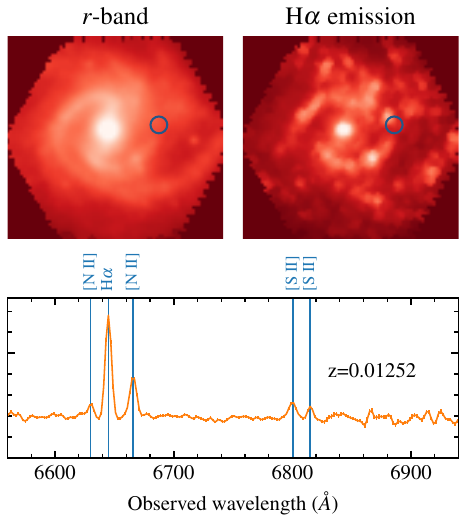}
    \caption{
        The top panels correspond to reconstructed $r$-band and H$\alpha$ images of NGC~6004, the host galaxy of SN\,2024cld, generated from the CALIFA/PMAS merged data cubes. The dark blue circle shows the 1\arcsec radius aperture centred on the explosion site of SN\,2024cld used to derive the explosion site spectrum. An H~II region is visible just off the edge of the explosion site, likely yielding some of the narrow line components seen in later-time spectra.
        The explosion site spectrum is plotted in the bottom panel, with relevant host lines used to fit for the redshift overplotted.
    }
    \label{fig:califa}
\end{figure}

NGC~6004 was observed as part of the Calar Alto Legacy Integral Field spectroscopy Area survey \citep[CALIFA; ][]{Sanchez2012,Husemann2013,Walcher2014}, conducted on the Calar Alto 3.5m telescope with the Potsdam MultiAperture Spectrophotometer \citep[PMAS/PPak; ][]{PMAS} instrument, using both the V500 and V1200 gratings. To assess the properties of the explosion site of SN\,2024cld, we extract a spectrum from an aperture of 1\arcsec radius, centred on the co-ordinates of SN\,2024cld, in order to measure the properties of the stellar population at the explosion site. Figure~\ref{fig:califa} shows $r/\text{H}\alpha$ maps of the host galaxy, the extraction aperture, and derived spectrum. The explosion site of SN\,2024cld lies close to a small H~II region.
Through joint fits to the H$\alpha$, [N~II], and [S~II] host lines, we estimate the redshift of the explosion site of SN\,2024cld to be $z=0.01252\pm0.00001$, corresponding to a velocity of \qty{+3750}{\kilo\meter\per\second}. This value is in contrast to the fiducial redshift of the galaxy nucleus, $z=0.01277(6)$~\citep{Springob2005}. This difference in redshift between host nucleus and explosion site (amounting to \qty{\approx-75}{\kilo\meter \per \second}) considering the radial offset is fully consistent with the spatially-resolved velocity maps of \citet{Barrera-Ballesteros2014}.

We  estimate the metallicity at the explosion site with the \citet{Dopita2016} relation using the fluxes of the H$\alpha$, [N~II], and [S~II] emission lines, and find a metallicity of $12+\log{\left(\mathrm{O/H}\right)} = 8.85\pm0.12$. As a cross-check, we also compute the N2 ($\mathrm{[N II]\,\lambda6584}$/ $\mathrm{H\alpha\,\lambda6563}$) index, using the calibration of \citet{Pettini2004}. The measured N2 index is $-0.37\pm0.02$, yielding a metallicity $12+\log{\left(\mathrm{O/H}\right)} = 8.69\pm0.18$. In both cases, the uncertainty is entirely dominated by the systematic component arising from the dispersion of the calibration relations used. Both methods yield results in good agreement with each other. This suggests that the explosion occurred in a region with metallicity consistent with solar (or even marginally super-solar) abundance ($8.66\pm0.05$; \citealt{Asplund2004}), perhaps unsurprisingly given the host galaxy. This value will be placed into context with the other 98S-like SNe in Section~\ref{sec:discussion}.

We estimate the host galaxy extinction at the explosion site using the equivalent width of the Na~I~D lines in our highest-quality, medium resolution ($R\sim7000$) WiFeS spectrum taken at +25 days. The combined equivalent width of the Na~I~D doublet is $0.78\AA$, which corresponds to a colour excess of $E_{(B-V)}=0.0105$ using the relation of \citet{Poznanski2012}. This is also corroborated by fits to low-resolution spectra that yield similar estimates. For all further analyses, we combine this with the Galactic extinction ($E_{(B-V)} = 0.1076$) to yield a combined extinction of $E_{(B-V)}=0.1181$, or $A_V\approx0.37$, assuming a Milky Way $R_V=3.1$ in absence of a measured value. 

\section{Photometry} \label{sec:photometry}
\begin{figure*}
    \centering
    \includegraphics[width=\linewidth]{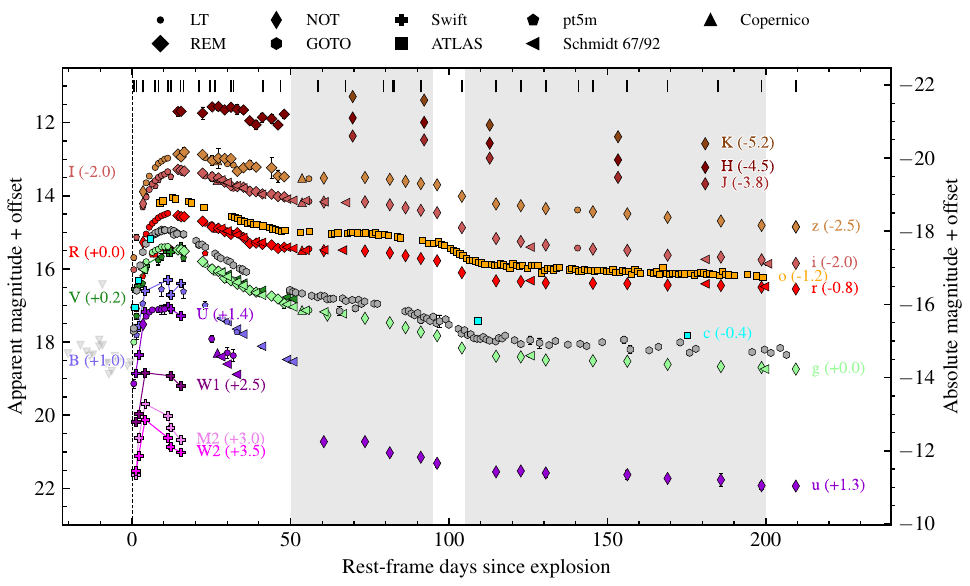}
    \caption{
        Host-subtracted photometry of SN\,2024cld, corrected for Galactic and host reddening.
        The light curve shows three key stages: a rapid rise to peak in $\sim$ 10 days, a plateau phase between 50-100 days post-explosion, and then a second plateau beginning around 110 days post-explosion.
        Non-detections are plotted as downwards-facing triangles.
        The explosion date is estimated from joint fitting of survey photometry (see Section~\ref{sec:explosion-fit}). Vertical tick-marks indicate the epochs on which spectroscopy was obtained. Error bars correspond to the $1\sigma$ photometric uncertainties. The two shaded regions mark the first and second plateaus referred to in the text.
    }
    \label{fig:lightcurves}
\end{figure*}

The collated light curve of SN\,2024cld is shown in Figure~\ref{fig:lightcurves}.
The light curve depicts an overall long-lived transient, at least compared to more typical SNe~II, with two clear plateaus in the light curve -- one starting around \qty{\approx50}{\day}, and another later one \qty{\approx110}{\day} post-explosion declining around 0.2 mag/100\,d in $r$-band, extending out to the final data obtained prior to SN\,2024cld becoming Sun-constrained around day 220. In the following subsections, we analyse the obtained photometry in greater detail.

\subsection{Explosion time and peak} \label{sec:explosion-fit}
\begin{figure}
    \centering
    \includegraphics[width=\linewidth]{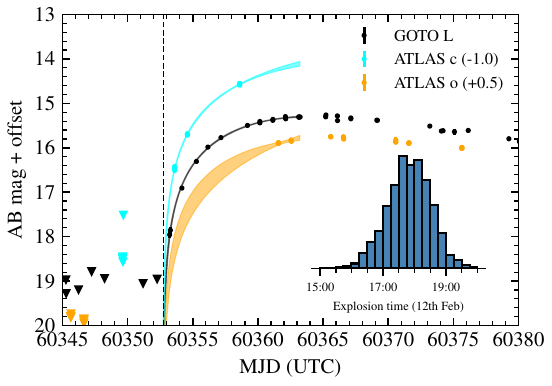}
    \caption{Pre-peak detections and upper limits for SN\,2024cld, overplotted with the model light curve for each survey used to infer the explosion date, which is marked with the dashed vertical line. Shaded regions correspond to the 1$\sigma$ confidence interval, estimated from posterior draws from the model. The triangles represent 5$\sigma$ non-detections. Offsets (given in the legend) are applied to the ATLAS photometry to enhance visibility. The inset histogram shows samples of the $t_\mathrm{expl}$ parameter to emphasise the tight constraints possible with generative modelling, plotted as UTC time on 2024 February 12.}
    \label{fig:explosion-fit}
\end{figure}

We estimate the explosion time of SN\,2024cld through jointly fitting the GOTO and ATLAS early-time light-curves with a simple hierarchical Bayesian model, similar to that of \citet{Miller2020} but without the power-law assumptions, to account for the clear interaction-dominated early phases. We assume the flux $f$ of each (early-time) light curve can be modelled with the piecewise expression
\begin{equation}
    f(\phi, B, a_0, a_1) = \begin{cases} 
          B & \phi < 0\\
          B + a_0 \phi + a_1 \phi^2 & \phi \geq 0 \\
       \end{cases}
\end{equation}
where $\phi$ is the modified time  since explosion, $t - t_\mathrm{expl}$, $B$ is a baseline correction term, and $a_i$ is the $i^{th}$ polynomial coefficient in $\phi$.  In the construction of the joint model, the explosion time ($t_\mathrm{expl}$) is the same for each light curve, yet the rise rate $a_{i,0}$ and change of rise rate $a_{i,1}$ vary per-filter. For numerical conditioning and correct error scaling we also assume a per-filter error rescaling $f_i$, and a per-filter baseline $B_{i}$ to allow for potential offsets in template fluxes, for example. We also only fit up to the point of maximum light in GOTO $L$ band, to ensure that only the early rise influences the explosion time estimate. Assuming strictly Gaussian errors in flux measurements, this yields the model likelihood
\begin{equation}
    \mathcal{L}(y_i | \theta) = \prod_{i=1}^{N} \frac{1}{\sqrt{2\pi \sigma_i^{\prime\,2}}} \exp\left(-\frac{(y_i - \mu_i)^2}{2\sigma_i^{\prime\,2}}\right)
\end{equation}
The prior distributions for all model parameters are assumed to be normal distributions to avoid imposing unnecessary inductive biases, excluding $t_\mathrm{expl}$ which is assumed to be uniform with lower and upper bounds corresponding to the last GOTO non-detection (MJD 60,352.0) and first GOTO detection (MJD 60,353.23), and the rise rate, which is constrained to be positive-definite by imposing a normal prior on the logarithm of the rise rate. We are  comfortably in the regime whereas the likelihood dominates the prior, and the other model parameters are considered nuisance. The model is constructed in JAX~\citep{jax2018github}, specifically the \texttt{numpyro} probabilistic programming language \citep{Phan2019}, and the posterior distribution is efficiently sampled with the No U-Turn Sampler~\citep{Hoffman2011} until the split R-hat~\citep{Vehtari2021} statistic is unity ($\sim$3000 warm-up steps). 
Figure~\ref{fig:explosion-fit} displays posterior draws of the joint light-curve model to validate the goodness of fit, alongside sampled explosion times from the posterior distribution.
We adopt the median of the joint model posterior as the inferred explosion epoch, with uncertainties computed from the \nth{16} and \nth{84} percentiles minus the median respectively -- yielding MJD $60,352.74 \pm 0.03$ (2024 February 12 17:45 UT). We here emphasise that this explosion epoch places the GOTO discovery just \qty{11.6\pm0.7}{\hour} post-explosion, and the first spectrum at \qty{12.4}{\hour}.

The peak absolute magnitude and time of maximum light are estimated through fitting polynomials (in the Chebyshev basis, for numerical stability) to the early-time light-curve, with uncertainties on each quantity derived from bootstrapping. Using the $g$-band, we find a peak magnitude of $m_g = 15.39$ mag, corresponding to a peak absolute magnitude of $M_g=-17.58$ mag at MJD 60,368.32 (\qty{14.65}{\day} post-explosion), incurring a systematic error of 0.64 mag from the distance modulus. The available photometry shows a trend towards longer rise times in redder filters. The rise is towards the longer end of the ordinary SNe~II presented in e.g.~\citet{Gall2015,Gonzalez-Gaitan2015}, with a fairly typical peak absolute magnitude. \citet{Gonzalez-Gaitan2015} find a time of $7.5\pm0.3$\,d for the majority of their sample -- with a marked tail of \say{long-risers}, which are compatible with the rise time of SN\,2024cld. The extended rise timescale is not especially surprising given the strong CSM interaction present, but still peaks quicker than SNe~IIn~\citep[e.g.][]{Taddia2013,Nyholm2020,Ransome2025}.

\subsection{Bolometric light curve}
\begin{figure}
    \centering
    \includegraphics[width=\linewidth]{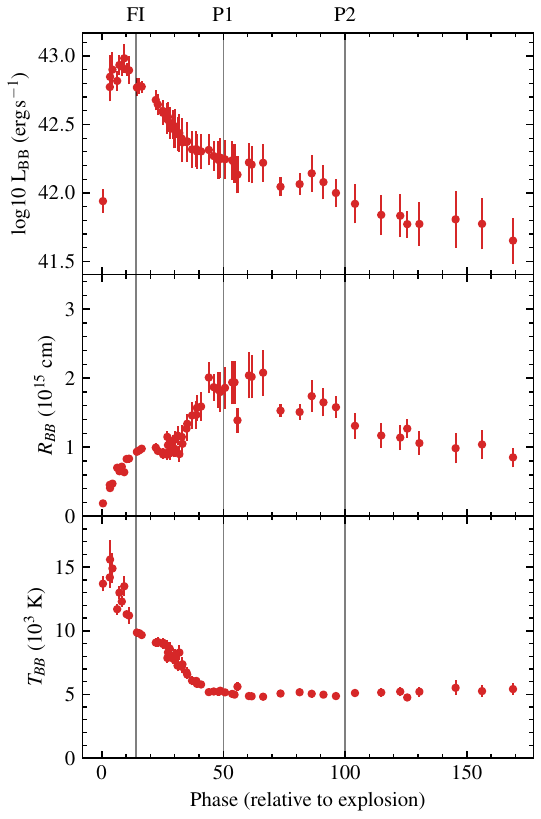}
    \caption{\textbf{Top:} pseudo-bolometric light curve of SN\,2024cld. \textbf{Middle:} Evolution of the blackbody (BB) radius of SN\,2024cld. \textbf{Bottom:} Temperature evolution of SN\,2024cld. The vertical lines across all panels show the end of the flash-ionised phase (FI), (approximate) start of plateau 1 (P1), and (approximate) start of plateau 2 (P2) to guide the eye. In all panels, the error bars correspond to the 1$\sigma$ uncertainty.}
    \label{fig:bolometric_lc}
\end{figure}

We construct the pseudo-bolometric light curve of SN\,2024cld via blackbody (BB) fits to the extinction-corrected multi-color photometry, interpolated/extrapolated to each $L$-band epoch as appropriate. Specifically, the \texttt{superbol} code described in \citet{Nicholl2018} was used. As we only have \textit{Swift} UV data until day $\sim$ 25, we assume constant color at later epochs. The resulting pseudo-bolometric light-curve and associated blackbody parameters are given in Figure~\ref{fig:bolometric_lc}.
SN\,2024cld reaches a peak pseudo-bolometric luminosity of $\log L=43.02\pm0.09~\mathrm{erg\,s^{-1}}$, declining quickly after this until the first plateau.
The integrated pseudo-bolometric luminosity over the first 100 days amounts to $2.4\times10^{49}\,\mathrm{erg}$.
The middle panel of Figure~\ref{fig:bolometric_lc} displays the evolution of the BB radius. At early times ($\lesssim14$\,\unit{\day}) the BB radius increases linearly, stalling as the ejecta sweeps out the flash-ionised material, and then increasing during the first light-curve plateau ($\approx50$\,d). Through fitting a straight line to the early-time (\qty{\leq 7}{\day}) BB radius and appropriately sampling uncertainties in explosion time, the BB radius at t=\qty{0}{\day} corresponds to \qty{1.5e14}{\centi\meter}, or $2100\pm200\,R_\odot$. This is larger than the typical radii of RSG stars, $\sim1500\,R_\odot$ at max~\citep[e.g.][]{Levesque2010},  and is further indicative of a highly-extended envelope surrounding the progenitor -- corroborating the extended rise time observed. 
The bottom panel of Figure~\ref{fig:bolometric_lc} charts the temperature evolution of SN\,2024cld. The BB temperature rises initially for the first 5 days (further supporting the evolving ionisation state inferred for the CSM, see Section~\ref{sec:spectroscopy}), peaking around \qty{15000}{\kelvin}, then begins to decline as typical of CCSNe. This decline levels out and remains constant around \qty{5000}{\kelvin} out to our last photometric measurements.

\subsection{Constraints on pre-explosion outbursts} \label{sec:pre-explosion}
\begin{figure}
    \centering
    \includegraphics[width=\linewidth]{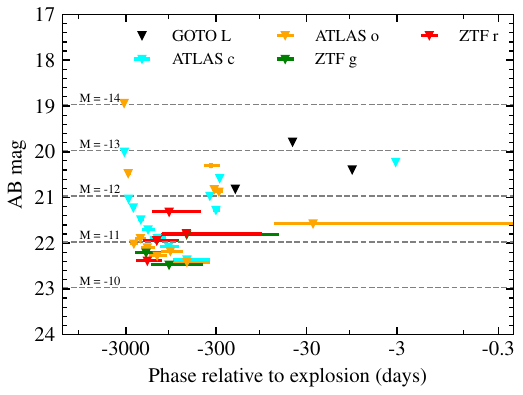}
    \caption{Photometric limits on pre-explosion precursor outbursts/variability for SN\,2024cld. Horizontal errorbars denote the (half-) width of the photometric bins -- with the seasonal binning used to condense the number of points present. Dashed lines show the corresponding absolute magnitude limits reached.
    }
    \label{fig:pre-explosion}
\end{figure}

Given the extended envelope of material we infer from the early-time photometry, and the timescale for this being ejected (see Section~\ref{sec:spectroscopy}), we conduct an archival search for any pre-explosion variability or precursor events using all publicly-available imaging. Figure~\ref{fig:pre-explosion} illustrates upper limits on pre-explosion activity in SN\,2024cld, constructed from stacking a mixture of ZTF, ATLAS, and GOTO photometry. These are combined in flux-space, using $\sigma$-clipping to remove outlying points, baseline correction, combining measurements with inverse-variance weighting, and \say{seasonal} binning to group detections to improve flux limits.
There are no compelling detections of a precursor outburst in any survey stacks. There is a single ATLAS detection in our seasonal stacks corresponding to MJD 60,089, but we attribute this to a period of poorer data quality from inspection of the frames, rather than a genuine source.

Based on the non-detections, we compute 5$\sigma$ upper-limits to any pre-explosion activity, and display these in Figure~\ref{fig:pre-explosion}, alongside corresponding absolute magnitude limits based on the distance modulus of NGC 6004. The data broadly rule out precursors down to $M=-11$ mag \qty{\geq100}{\day} prior to explosion, i.e. comparable to the pre-explosion outburst of SN\,2020tlf~\citep{Jacobson-Galan2022} and SNe~IIn~\citep{Fraser2013,Reguitti2024}, but likely insufficient to probe further down the precursor luminosity function, which is believed to be steep, favouring fainter events. The data in hand, however, do disfavour events with absolute magnitudes similar to these literature examples. Nevertheless, absence of detections does not imply an absence of progenitor outbursts or variability, especially in the context of an optically thick CSM shrouding the progenitor during any episodes. We discuss the implications of this in more detail in Section~\ref{sec:discussion}.

\section{Spectroscopy} \label{sec:spectroscopy}
\begin{figure*}
    \centering
    \includegraphics[width=0.95\linewidth]{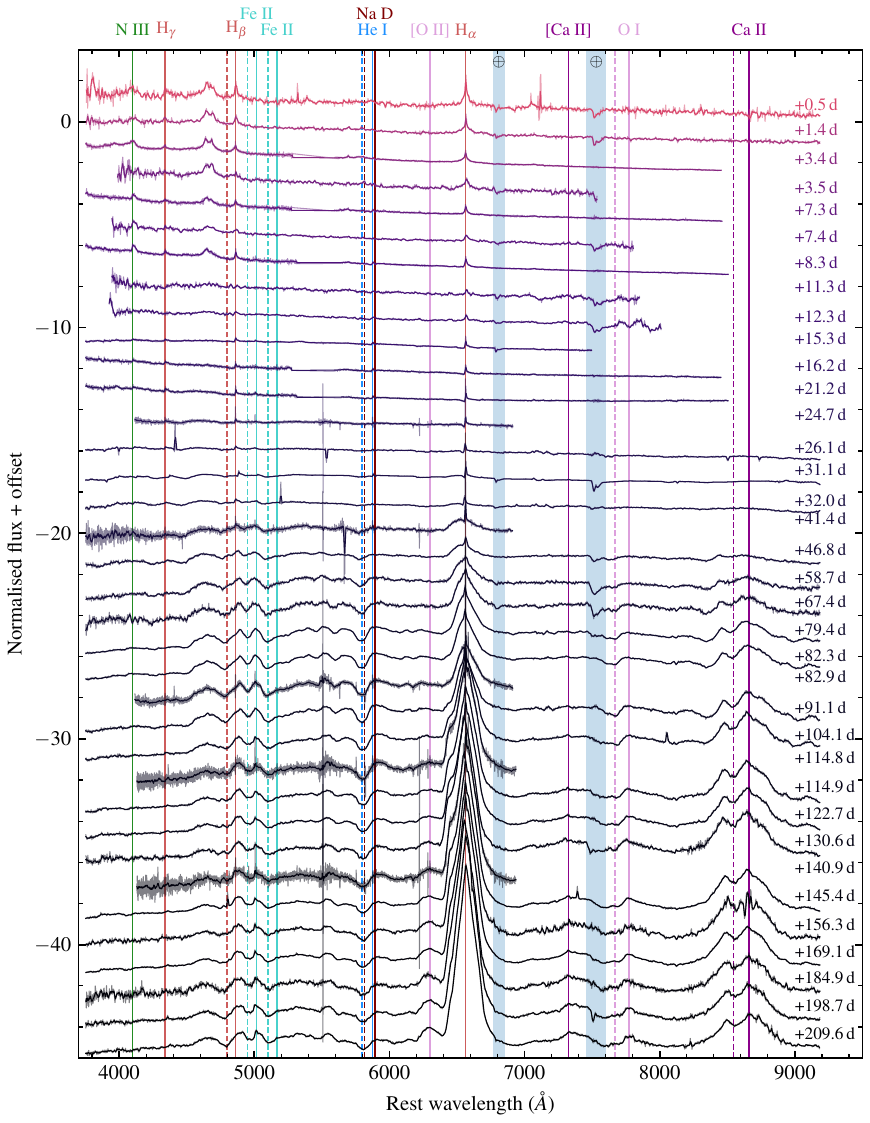}
    \caption{Full spectral series for SN\,2024cld, recalibrated to photometry and corrected for Galactic and host reddening. The two shaded regions correspond to major telluric absorption features. The solid lines correspond to rest wavelengths, with the dashed lines showing a \qty{-5000}{\kilo\meter\per\second} velocity offset, consistent with the bulk of the ejecta at late time. NOT medium-resolution spectra are excluded as they only span a small range of wavelengths, and deferred to Figure~\ref{fig:highres_spectral_series}.}
    \label{fig:full-spectral-series}
\end{figure*}

We present the full spectral series of SN\,2024cld in Figure~\ref{fig:full-spectral-series}, and in the following subsections break down the spectral evolution of SN\,2024cld into a number of key phases -- making comparisons to relevant literature objects and models, tracing the ejecta kinematics, and studying in detail the extensive dataset of medium-resolution spectroscopy obtained. Spectra for a number of comparison objects are retrieved via WISeREP\footnote{\url{https://wiserep.org}}~\citep{Yaron2012}, with the references to specific objects given in the text where appropriate.

\subsection{Early time: flash-ionised phase}
\begin{figure}
    \centering
    \includegraphics[width=\linewidth]{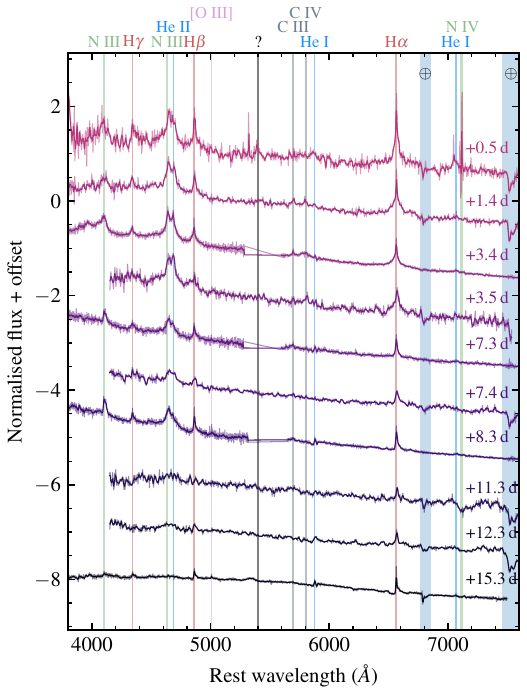}
    \caption{
        Early-time spectral series for SN\,2024cld, truncated to the first 15 days (roughly coincident with the disappearance of the flash-ionisation features). The spectra prior to +3.5d show a number of high-ionisation lines (C~III, N~IV) that quickly dissipate, leaving only H and He in latter spectra.
        Phases are relative to explosion time. Spectra have been de-reddened and recalibrated to contemporaneous photometry. The shaded regions correspond to major telluric absorption features.
    }
    \label{fig:spectralseries-earlyzoom}
\end{figure}

\begin{figure}
    \centering
    \includegraphics[width=\linewidth]{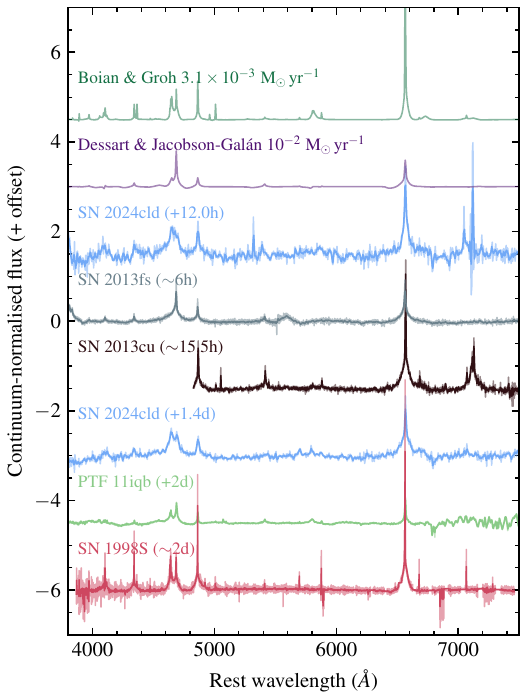}
    \caption{
        Continuum-normalised early-time spectra of flash-ionised SNe. 
        The \citet{Boian2020} and \citet{Dessart2023} models are convolved with a Gaussian kernel to approximately match the spectral resolution of the observations.
    }
    \label{fig:flash-features-compare}
\end{figure}

Figure~\ref{fig:spectralseries-earlyzoom} zooms in on the first $\sim$ 2 weeks of spectral evolution of SN\,2024cld, alongside line identifications for the early flash-ionisation phase.
The early flash-ionised spectra are dominated by H and He, showing the broad Lorentzian profiles typical of electron scattering. In both the \qty{+1.4}{\day} INT and \qty{+3.4}{\day} NOT spectra, higher ionisation species such as C~III, C~IV, and N~IV are clearly detected. Although the initial INT classification spectrum has lower S/N, the evolution between \qty{+1.4}{\day}  and \qty{+3.4}{\day}  suggests they were likely present at a weak level already in this spectrum compared to the clearly-present H and He~II. These higher-ionisation lines also notably broaden out with time. To further place these features in context, 
in Figure~\ref{fig:flash-features-compare} we compare SN\,2024cld against a number of comparison objects that show flash-ionisation.
These include the two prototypical flash-ionised SNe~II with very early spectra, SN\,2013fs~\citep{Gal-Yam2014} and iPTF13dqy/SN\,2013cu~\citep{Yaron2017}, the early-time spectra of SN\,1998S~\citep{Shivvers2015} and PTF~11iqb~\citep{Smith2015}. We also overplot the best-matching models from the grids of \citet{Boian2020} and \citet{Dessart2023} to aid line identification, though caution that the underlying assumptions are likely not well-founded in the specific case of SN\,2024cld, given the aspherical CSM that we infer in latter sections. SN\,2024cld sits comfortably between SN\,2013fs and SN\,2013cu in terms of phase, and shows a close match in terms of intensity and profile of H/He features. The detailed agreement of ions is more complex however, owing to signal-to-noise ratio - though SN\,2013fs clearly lacks the He~I/N~IV complex at $\lambda7100\AA$ that SN\,2013cu and SN\,2024cld (\qty{12}{\hour}) share. 

The FI phase lasts longer in SN\,2024cld than is typical -- there is a broad bump feature in the +11.3 and \qty{+12.3}{\day}  spectra in the location of the He~II/N~III complex, that is entirely gone in the \qty{+15.4}{\day}  INT spectrum. To be conservative, we set the end of the flash-ionised phase to be mid-way between the \qty{+12.3}{\day}  and \qty{+15.2}{\day}  spectra, i.e. \qty{13.9\pm1.4}{\day}.  This duration of flash-ionisation is towards the longer-duration end of the sample of normal SNe~II presented in \citet{Bruch2023} (lying $3.2\,\sigma$ from the mean). Although some other classes of CSM-interacting transients have shown signficantly longer FI periods~\citep[e.g. ][]{Kangas2025}, SN\,2024cld has among the longest FI phase of any SN~II, consistent with the idea of a dense, extended CSM present from a prolonged mass loss period driving longer rise times. The BB radius at the end of the FI phase (\qty{1e15}{\centi\meter}) is in good agreement with the expected radius for the material driving FI emission in other SNe~II from the sample of \citet{Jacobson-Galan2024}.

\subsection{Photospheric phase}
Beyond the early-time, the spectrum of SN\,2024cld broadly evolves as expected of a typical SN\,II, albeit with delayed emergence of the metal absorption lines, likely caused early-on by initially-high CSM interaction. The typical H$\alpha$ P Cygni profile seen in CCSNe is also markedly absent, with the absorption filled in entirely by (presumably) CSM interaction-driven emission~\citep[][]{Pessi2023}. This is in common with both SN\,1998S and PTF11iqb.

\begin{figure}
    \centering
    \includegraphics[width=\linewidth]{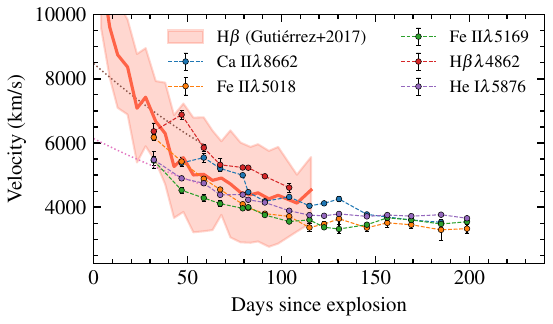}
    \caption{Evolution of ejecta velocities in SN\,2024cld through the first 200 days. Velocities correspond to the profile minima of the associated absorption, with error bars representing bootstrapped 1$\sigma$ uncertainties. The extrapolated velocities back to zero-phase in brown and pink dashed lines correspond to H$\beta$ and He~I, respectively, and illustrate the expected stratification in ejecta. The measured mean H$\beta$ velocities from the sample of \citet{Gutierrez2017}, alongside their standard deviations (shaded region) are overplotted for comparison.}
    \label{fig:ejecta_velocities}
\end{figure}

To further underscore the normality of the underlying explosion in SN\,2024cld, we investigate the ejecta kinematics through measurement of the ejecta-driven absorption lines throughout the lifetime of the transient. Owing to the complex profiles of the ejecta lines, we estimate the profile minima by fitting Chebyshev polynomials to selected spectral ranges around each line of interest, using primarily the NOT gr4 spectra owing to their high signal-to-noise. Uncertainties are estimated by bootstrap resampling to account for data-driven uncertainty in the location of the profile minimum. Prior to \qty{\sim30}{\day} the metal absorption features are not visible. Beyond \qty{100}{\day} we do not attempt to fit for the H$\beta$ velocity, owing to major contamination from Fe lines. Similarly, some later-time Ca~II NIR line measurements are excluded owning to residual fringing and/or telluric absorption.
In Figure~\ref{fig:ejecta_velocities} we plot the results of this fitting procedure to compare and contrast the velocities of the different species in the ejecta, as well as estimate initial ejecta velocities.
The best-determined ejecta velocity arises from the Fe~II$\lambda$5169 line, typically taken as a proxy for the photospheric velocity~\citep[e.g.][]{Schmutz1990,Dessart2005}. This, alongside all other line velocities, shows very typical evolution compared to the sample of \cite{Gutierrez2017}, further underscoring that SN\,2024cld is indeed a typical SN\,II, with its unusual properties arising from CSM interaction.

\begin{figure}
    \centering
    \includegraphics[width=\linewidth]{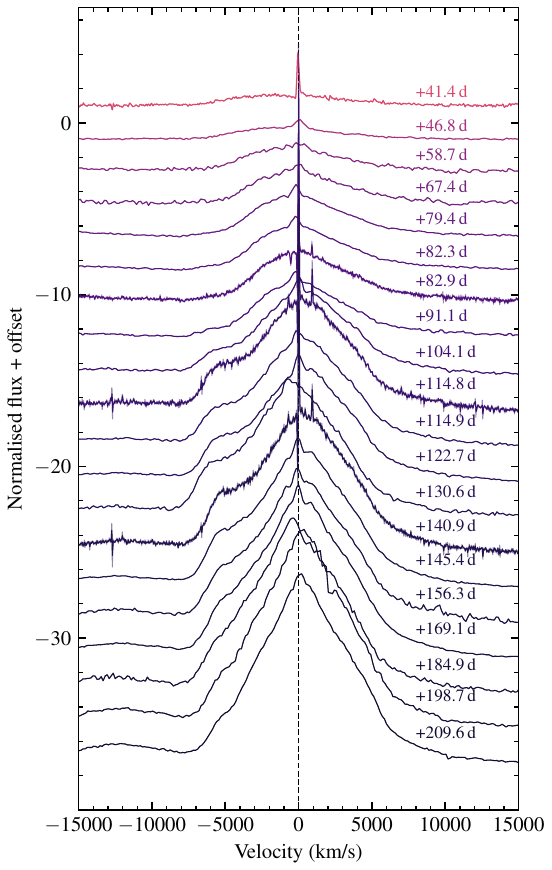}
    \caption{Evolution of the broad H$\alpha$ profile in SN\,2024cld. A strong blue shoulder and weaker red ledge are present.}
    \label{fig:halpha_evol}
\end{figure}

To investigate what drives the early-time rise to peak  (noted in Section~\ref{sec:photometry}) in the photospheric radius, we estimate the ejecta velocity close to time of explosion. We fit the obtained velocities of H$\beta$ and He~I with low-order polynomials to estimate the ejecta velocities at day zero and obtain \qty{8700}{\kilo\meter\per\second} and \qty{6700}{\kilo\meter\per\second}, respectively. These are formally lower limits, given that the early-time evolution is not well-described by a linear decrease. Following the end of the first plateau at \qty{\sim100}{\day}, the broad H$\alpha$ profile observed in SN\,2024cld undergoes two key changes -- the emergence of a strong blue-shifted emission component with a velocity -\SIrange{5000}{6000}{\kilo\meter\per\second}, and a weaker red \say{ledge} feature at similar velocity. Figure~\ref{fig:halpha_evol} illustrates the evolution of the H$\alpha$ profile over time.
We perform a profile decomposition to investigate the underlying components, by fitting a number of Gaussians to the H$\alpha$ emission. The emission seen in SN\,2024cld is well-described by four Gaussians: a narrow emission component at a rest wavelength corresponding to the pre-shock CSM emission/host light, a broad emission component close to rest wavelength arising from ejecta, and blue and red components offset \qty{\sim6000}{\kilo\meter\per\second} from rest wavelength, corresponding to high-velocity CSM components. We also include a component to account for the emergence of nebular oxygen at late times, fixed to the rest wavelength of \qty{6300}{\angstrom}. Note that given we use the lower resolution spectra for this analysis, we do not include a detailed description of the narrow pre-shock CSM emission with the P Cygni absorption -- instead deferring this to Section~\ref{sec:highres}. All lines are fit simultaneously to correctly propagate covariances. Example fits are shown in Figure~\ref{fig:halpha_fits}, with individual components isolated.

Prior to $\approx90$d, the profile is largely ejecta-dominated, with a slight excess visible on the red wing, generated by a broad, high-velocity component with full width at half maximum (FWHM) \qty{\approx7000}{\kilo\meter\per\second}, at typical velocity offset \qty{\approx6000}{\kilo\meter\per\second}. After $\approx90$d, a blue component in H$\alpha$ emerges at a velocity offset \qty{\approx5500}{\kilo\meter\per\second}, driving the emergence of a marked \say{shoulder} in the profile on a similar timescale.  Both red and blue components remain until our final spectrum - though with diminished strength owing to the decrease in continuum flux and strengthening of the ejecta-driven H$\alpha$ emission. The red component typically has a 2--3$\times$ larger FWHM than the blue component throughout the spectral series. There is no real evidence for dust formation in the NIR photometry (see Figure~\ref{fig:bb_fits}) of SN\,2024cld at these epochs, which could modulate the wings of the Balmer lines, thus we favour the explanation that these changes are indeed related to CSM components from an asymmetric configuration.

Outside of the H$\alpha$ profile, the late-time spectral features of SN\,2024cld are broadly comparable to typical CCSNe, showing emergence of nebular emission, but with a few deviations. The [Ca~II]$\lambda\lambda$7291, 7324 and [O~I]$\lambda$7774 lines, typical of late-time spectra of SNe~II~\citep[e.g. ][]{Dessart2020} are notably absent -- whereas the Ca~II NIR triplet remains strong throughout the final \qty{\sim100}{\day} of our spectral sequence. 
The relatively weak strength of the forbidden emission lines is best explained by the elevated CSM density present~\citep[e.g. ][]{Filippenko1997}. The higher densities in the CSM/ejecta-CSM boundary regions exceed those at which forbidden transitions can occur, and thus lead to suppression of [Ca~II] relative to Ca~II, for example. This is relatively common among the class of interacting SNe, with strong IIns~\citep[e.g.][]{Taddia2013} showing an almost complete suppression, with more weakly-interacting SNe~II~\citep[e.g.][]{Nagao2025} showing weakened forbidden lines. We defer a comparison of forbidden lines in SN\,2024cld to the other 98S-like SNe, along with more detailed spectral comparisons to Section~\ref{sec:discussion}. 
    There is no strong evidence for the Ca~II triplet and O~I line showing the same asymmetries as H$\alpha$, particularly around the phase of the emergence of the blue shoulder (\SI{+100}{\day}), when it might be expected to manifest most strongly.

At the end of our spectral series, SN\,2024cld is still photospheric, with strong Fe~II absorption features present. This is further evidence for CSM interaction heating the ejecta, especially in light of the  late-time interaction-driven luminosity, which endures for more than \qty{100}{\day} longer than seen in the sample of \cite{Anderson2014}. This makes modelling of the progenitor star challenging at this early phase, with spectroscopy at a later time, when the interaction has ceased and the ejecta have transitioned to being optically thin, crucial to make further headway in confirming the progenitor of SN\,2024cld in detail. 

\subsection{Medium-resolution spectroscopy}
\label{sec:highres}
\begin{figure}
    \centering
    \includegraphics[width=\linewidth]{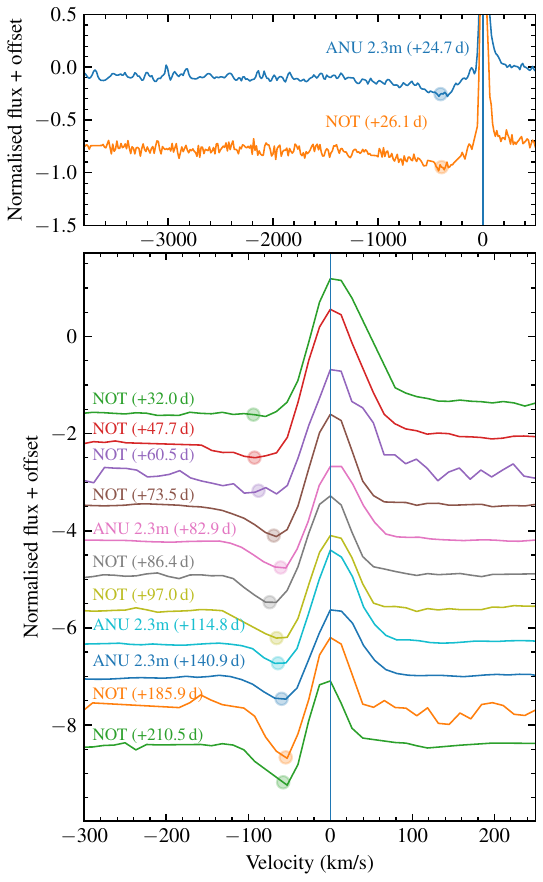}
    \caption{Zoom-in on the narrow H$\alpha$ component present in medium-resolution data of SN\,2024cld, with spectra shifted to match the H emission. The top two spectra are plotted with a expanded scale to emphasise the broad absorption present. There is a marked transition around \qty{+30}{\day} post-explosion, attributed to the ejecta sweeping past dense, eruptive mass loss, into a more rarified wind-dominated CSM. The shaded circle corresponds to the estimated profile minimum for each spectrum.}
    \label{fig:highres_spectral_series}
\end{figure}

\begin{figure}
    \centering
    \includegraphics[width=\linewidth]{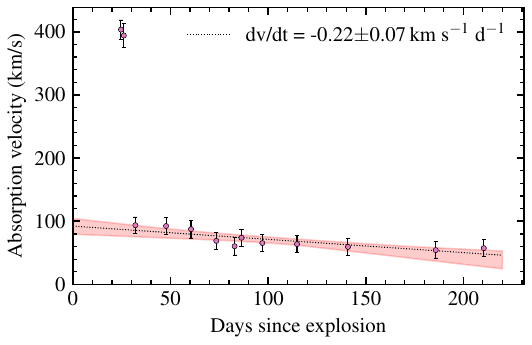}
    \caption{Measured H$\alpha$ absorption velocities as a function of phase, based on medium-resolution NOT and ANU2.3m data. Measurements and their 1$\sigma$ uncertainties are derived from bootstrapped Chebyshev polynomial fits to the line profile. Overplotted is a linear fit to the post-transition velocities to illustrate the regression to the terminal wind velocity, alongside 1$\sigma$ bootstrapped confidence interval.}
    \label{fig:absorption_vel}
\end{figure}

Given the presence of narrow lines even after the initial flash-ionised stages, we triggered a sequence of higher spectral resolution observations beginning \qty{+24}{\day} post-explosion, using NOT/ALFOSC and ANU2.3m/WiFeS in their highest-resolution configurations. These spectra are shown in Figure~\ref{fig:highres_spectral_series}.
Our best NOT/ALFOSC gr17 observations taken with the \qty{0.5}{\arcsec} slit have typical spectral resolution\footnote{Sampled at approx. 0.2$\AA$/pix} $R\approx9000$ measured from the FWHM of unblended sky lines, corresponding to an instrumental broadening of \qty{34}{\kilo\meter\per\second}.
Throughout the post-flash evolution (\qty{\geq15}{\day}), we detect a narrow (FWHM=\qty{65}{\kilo\meter\per\second}), corrected for instrumental broadening) H$\alpha$ emission component consistent with the host galaxy rest frame, initially attributed to the nearby H~II region (see Figure~\ref{fig:califa}). However, a lack of accompanying [N~II] lines, and more crucially, a deep P~Cygni profile, suggest this is intrinsic to the SN itself. Most remarkably, this P~Cygni profile shows evolution throughout the lifetime of the transient, illustrated in Figure~\ref{fig:highres_spectral_series}.
At early times, the absorption is broader (\qty{\approx200}{\kilo\meter\per\second} FWHM) and has a profile minimum around \qty{500}{\kilo\meter\per\second}. A sharp transition occurs around day 30, with the absorption velocity dropping to \qty{\approx100}{\kilo\meter\per\second}, and the profile width decreasing dramatically. We cannot rule out the possibility of further narrow absorptions at this phase owing to the signal-to-noise ratio and resolution of the data. We associate this with an abrupt change in the CSM structure swept by the ejecta around this time -- with early-time velocities corresponding to intense mass loss from e.g. binary interaction or ejected material from an eruption, and later-time velocities corresponding to an unperturbed wind-driven CSM. For such a sharp transition to occur implies a change in CSM on length scales of $\leq$20 AU. The transition from broad-high velocity to narrow-lower velocity is further consistent with the complex nature of eruptive (or binary) mass loss. We explore in more depth the implications of this result in Section~\ref{sec:discussion}, but for now characterise the CSM velocity in more detail.
To more systematically measure the evolution of this feature, we fit Chebyshev polynomials to the absorption to estimate the profile minimum. Velocities are referenced to the strong H$\alpha$ emission to mitigate any wavelength calibration offsets (thought to be minimal owing to an absolute calibration to sky-lines). These velocities are presented in Figure~\ref{fig:absorption_vel}.
There is an evident decline post-transition, which we fit with a straight line -- valid given we are still seeing absorption from comparatively close in to the progenitor, which has not yet reached terminal velocity. There is a clear decrease in velocity over time, \qty{0.22\pm0.08}{\kilo\meter\per\second\per\day}, which can be attributed to the ejecta sweeping up faster material (i.e. closer to the SN). This should decrease towards the terminal wind velocity~\citep[e.g.][]{Kee2021} over time if this component is indeed associated with the progenitor wind, as the effects of radiative acceleration decrease with greater radius -- with decreasing wind velocities placing stronger and stronger constraints on this. The final value obtained prior to SN\,2024cld becoming Sun-constrained, \qty{57\pm13}{\kilo\meter\per\second}, is above typical RSG wind velocities (\SIrange{10}{30}{\kilo\meter\per\second}), though broadly consistent within the uncertainties. This profile will likely disappear prior to reaching the terminal wind velocity, as the CSM density continues to decrease towards the point it can no longer drive line emission.  Such narrow absorptions have been seen in a number of prior SNe~II from the literature \citep[e.g. ][]{Shreshtha2024,Andrews2025,Tartaglia2025}, but few have shown the longevity of the emission seen in SN\,2024cld -- the strongly-interacting SN\,2017hcc being a notable exception~\citep{Smith2020_2017hcc}.

\section{Polarimetry} \label{sec:polarimetry}
\begin{figure*}
    \centering
    \includegraphics[width=\linewidth]{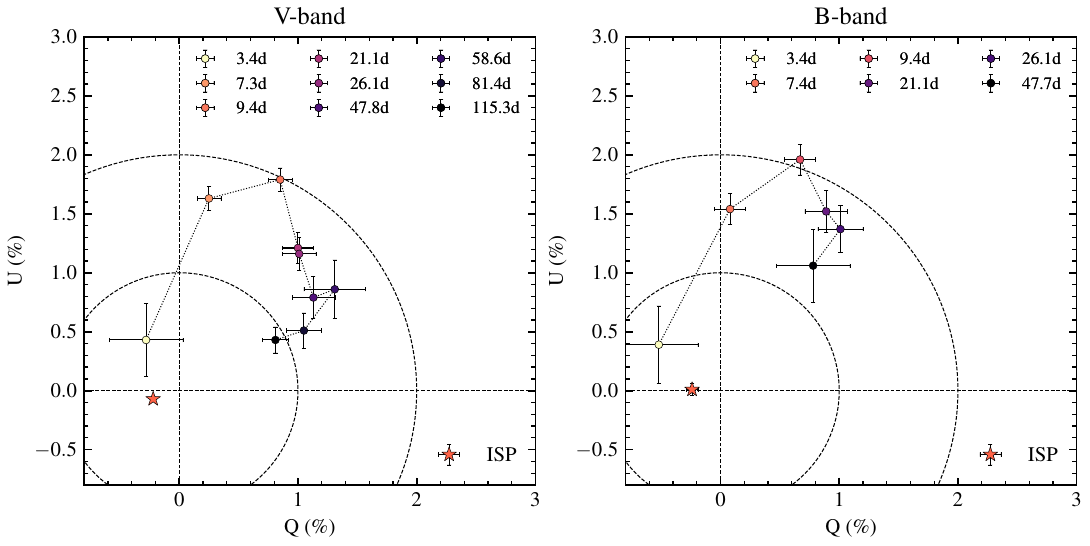}
    \caption{$B$ and $V$-band polarisation measurements for SN\,2024cld with NOT/ALFOSC, based on an aperture size of 2$\times$FWHM, and corrected for ISP via field stars. There is clear and complex polarimetric evolution throughout in both $B$ and $V$ bands, rising to $P\approx2\%$ around \qty{10}{\day} post-explosion, and showing marked rotation of \ang{60} from explosion to final epoch. The ISP (and associated uncertainty) is shown by the red marker.
    }
    \label{fig:pola_QUdiagram}
\end{figure*}

The optical polarimetry of SN\,2024cld samples the early and mid-time evolution of the transient well, ending around \qty{115}{\day} post-explosion owing to it becoming too faint for reliable observations with the NOT. For the earliest stages of evolution we have both $B$ and $V$-band observations, both tracing the continuum polarisation well due to avoiding the prominent H$\alpha$ line.
The earliest epoch required some additional calibration owing to a sub-optimal observing configuration, which we detail here prior to exploring the polarimetric evolution. 
The first epoch of polarimetry at \qty{+3.5}{\day} suffered from excess contamination of SN\,2024cld by the extraordinary beam of the host nucleus, which made a straightforward reduction challenging. We applied two approaches to remove the contribution of the host nucleus -- subtraction of the flux of the nucleus based on an empirical radial profile based on Chebyshev polynomials (approximately valid for the core, where the majority of the flux is located), and subtraction of the flux of the nucleus from the contaminated epochs based on measurement of latter uncontaminated epochs using the $e/o$ beam ratio. Both approaches are independent, and yet led to consistent results, thus we are confident that this first epoch is justified to include.
The interstellar polarisation (ISP) due to Galactic dust was estimated based on field stars to be $0.24 \pm 0.05$\% in $B$-band, and $0.23\pm0.02$\% in $V$-band. Given our inference of low host extinction from medium-resolution ANU~2.3m spectra (see Section~\ref{sec:hostglx}), the contribution to polarisation from the host (e.g. see \citealt{Serkowski1975}) is likely small, $\lesssim0.1\%$, at least compared to the $\sim2\%$ polarisation measured throughout the evolution of SN\,2024cld. The full polarimetric series is presented in Figure~\ref{fig:pola_QUdiagram}, corrected for ISP and all instrumental effects.
SN\,2024cld shows a number of notable features in polarimetric evolution. The earliest measurements in both $B$ and $V$ band are very low, broadly consistent with zero polarisation at the 1-2$\sigma$ level. Over the next week, the polarisation in both bands rises rapidly to $\approx$2\%, and rotates approximately $40\,\deg$. This is the maximal observed polarisation attained by SN\,2024cld throughout our observed epochs. After this, the polarisation declines slowly in both bands, and shows evidence for further rotation to a maximum of $\approx$50-60$^{\circ}$ with respect to the first measurements. Our final $V$-band epoch at \qty{115}{\day} post-explosion remains elevated at the $\sim$1\% level. Formally, these measurements constitute lower limits, given the unknown viewing angle adding projection effects. 

Prior polarimetry of 1998S-like SNe in the literature is largely limited to single epochs or sparsely-sampled imaging/spectro-polarimetry datasets. The detailed polarimetric observations of SN\,2024cld across a broad timespan presented here thus represent the most comprehensive dataset for any 98S-like SN in the literature, and offer a valuable opportunity to better understand the geometry of CSM in these objects, and thus constrain their progenitor systems better than previously possible. There are 3 epochs of spectropolarimetry publicly reported for SN\,1998S, presented in \citet{Leonard2000} and \citet{Wang2001}, obtained approximately +4, +27, and +59\,d post discovery respectively. The earliest epoch shows continuum polarisation of 1.6\% under the revised ISP of \citet{Wang2001}, before undergoing a rotation of $\approx36\deg$ in the second epoch 1 month post-discovery, and growing to 3\% in the final epoch with constant polarisation angle. A single epoch of spectropolarimetry~\citep{Bilinski2024} was obtained for PTF11iqb at +176\,d post-discovery, showing polarisation of $3.0\pm0.9\%$ at 5400Å (approximately $V$-band).
The observed polarimetric properties of SN\,2024cld are well-matched to these two objects -- showing an initially low polarisation ($\approx0.5\%$) at discovery, which grew to a peak of $\approx2\%$ at +10\,d. SN\,2024cld also shows the same rotation of $\approx40\,\deg$ as reported in SN\,1998S, corresponding to the change in geometry of the photosphere as the ejecta sweeps the inner CSM. At late time (+115\,d), SN\,2024cld shows $p\approx1\%$, consistent with the 3$\sigma$ detection of non-zero polarisation reported in PTF~11iqb. The overarching similarities in polarisation evolution between SN\,2024cld and these objects further underscores the common presence of asymmetric CSM, even if detailed agreement is harder to assess given sparse literature data. Such high levels of polarisation are also seen in the more strongly interacting SNe~IIn \citep[see e.g.][]{Bilinski2024}, which map well to the evolution seen in the more weakly-interacting SNe we focus on here, as expected given that asymmetric CSM is also often invoked to explain their properties \citep[e.g. ][]{Reynolds2025b}. In these objects however, it is clear that significantly higher CSM masses are involved given the higher luminosities, pointing towards different progenitors in the most extreme cases~\citep[e.g. ][]{Fransson2014_2010jl,Ofek2014_2010jl}. 

The polarimetric evolution SN\,2024cld shows excellent consistency with recent ordinary SNe~II with very early polarimetry -- albeit over a longer timescale, and with greater overall observed polarisation. Spectropolarimetry of SN\,2024ggi~\citep{Yang2025} shows rising polarisation from \qty{1.2}{\day} post-explosion, peaking at around \qty{2}{\day} ($p\approx0.5\%$) before declining, and showing a large-scale rotation in the dominant axis. A similar trend is seen in SN\,2023ixf~\citep{Shrestha2025}, with an initially high polarisation ($p\approx1\%$) from their first measurements around \qty{2}{\day} post-explosion, with further data showing both a rapid decline, and an sharp change in position angle between days 2 and 12. The qualitative evolution of SN\,2024cld shows a good match to both objects, with their polarisation evolution being attributed to aspherical CSM close-in to the explosion (and/or asymmetric ejecta). The slower rise to peak polarisation, smaller rotation, and continuous detection of polarisation out to \qty{+120}{\day} seen in SN\,2024cld further underscore the highly extended nature of the CSM in this system, as already inferred from spectroscopy. One notable deviation is further explained by the extended CSM present in SN\,2024cld: the presence of elevated polarisation on/after the H-recombination plateau. In ordinary SNe~IIP~\citep{Chornock2010,Nagao2024} this phase is marked by spherical symmetry, but in SN\,2024cld the extended, asymmetric CSM present (see Section~\ref{sec:spectroscopy}) must be driving an aspherical photospheric geometry even through this phase to explain the evolution seen.

\section{Discussion} \label{sec:discussion}
\subsection{Bulk properties in comparison to other 98S-likes}
\begin{figure}
    \centering
    \includegraphics[width=\linewidth]{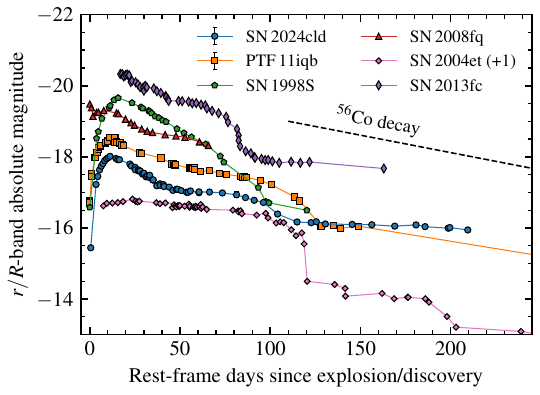}
    \caption{Comparison between the light curves of 98S-like SN\,2024cld (this paper), PTF11iqb, SN\,1998S, SN\,2008fq, SN\,2013fc, and normal SN\,II SN\,2004et. Absolute magnitudes are corrected for Galactic reddening, and host reddening where available. A moving average with a 1-week window is applied to the light curve of SN\,2013fc to reduce scatter. A representative $^{56}$Co decay assuming full gamma-ray trapping is overplotted to highlight the significant excess luminosity over radioactive decay.}
    \label{fig:lightcurve_comparison}
\end{figure}

\begin{figure}
    \centering
    \includegraphics[width=\linewidth]{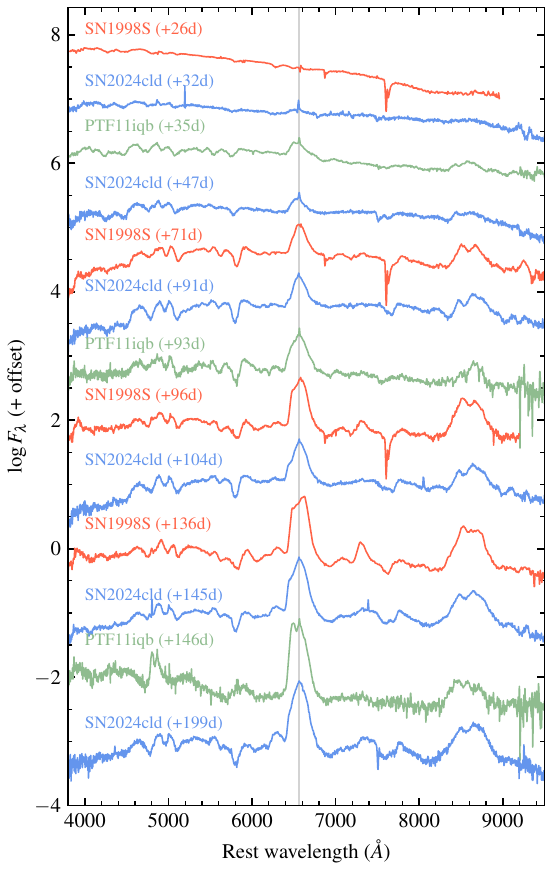}
    \caption{Spectral comparison between SN\,2024cld, PTF11iqb, and SN\,1998S. All spectra are corrected to their rest frame, and we apply corrections for host reddening.}
    \label{fig:spectral_comparison}
\end{figure}

In this section, we compare and constrast the observed properties of SN\,2024cld to the other 98S-like objects to infer where they differ in terms of CSM interaction and underlying explosion. For the purposes of comparison, we present in Figure~\ref{fig:lightcurve_comparison} the $r/R$-band light curves of a number of 98S-like transients from the literature: PTF11iqb~\citep{Smith2015}, SN\,1998S~\citep{Fassia2000,Poon2011}, SN\,2008fq~\citep{Taddia2013}, and SN\,2013fc~\citep{Kangas2016}. We also include the well-observed normal SN\,II SN\,2004et~\citep[e.g.][]{Maguire2010} for comparison. $r/R$-band is deliberately chosen to capture the late-time luminosity from H$\alpha$, which dominates the overall luminosity out to later times - with differences between the two filters being minimal as a result.
We also present a more detailed spectroscopic comparison in Figure~\ref{fig:spectral_comparison} with PTF11iqb and SN\,1998S, using spectra retrieved from WISeREP, originally presented in \citet{Smith2015} and \citet{Fassia2001, Leonard2000} respectively.

The early time light curve of SN\,2024cld shows a marginally extended rise compared to regular SNe~II, in common with PTF11iqb and SN\,1998S. The rise time is comparable to PTF11iqb \citep[$\approx$14\,d; ][]{Smith2015}, with both being quicker to peak than SN\,1998S~($\gtrsim16$\,d, with some uncertainty on the explosion epoch). Taken together, this hints towards the early-time interaction in SN\,2024cld being weaker than in the other 98S-likes -- further corroborated by the peak luminosity being comfortably lower than both. The initial photospheric radius at time of explosion, inferred from the bolometric light curve, implies a significantly puffed-up progenitor compared to typical RSGs -- with mechanisms for achieving this being suggested as envelope inflation via wave-driven heating~\citep{Quataert2012,Fuller2017}, pulsations~\citep{Yoon2010}, or late-stage nuclear burning~\citep{Morozova2020}. The CSM interaction is further evident in the early-time spectra, showing characteristic flash-ionisation signatures in common with many CCSNe, and indeed PTF11iqb itself. The narrow emission lines attributed to CSM interaction disappear around 14 days post-explosion, which corresponds with a stall in the photospheric radius. The photospheric temperature throughout this early phase is comparatively low, around \qty{15000}{\kelvin} at peak, and consistently decreasing. This inferred temperature is markedly lower than seen in other flash-ionised SNe at early times \citep[e.g.~][]{Yaron2012,Terreran2022}, but is broadly comparable to SN\,1998S~\citep{Fassia2000} at similar phase. The high-ionisation species are likely driven by ejecta-CSM interaction rather than the more classical flash-ionisation: the early rise in the photospheric radius has a velocity comparable to the inferred ejecta velocity (see Figure~\ref{fig:bolometric_lc}) and explains the extended interaction seen compared to the sample of \citet{Bruch2023}. From when the material causing the narrow flash lines is swept up by the advancing ejecta, we can infer the outer extent of this dense component based on the inferred ejecta velocity at early time (\qtyrange{8000}{10000}{\kilo\meter\per\second}) as approximately 60-80 AU ($\sim 10^{15}$\,cm) across. The observed polarisation degree also peaks on a similar timescale to when the flash lines disappear, suggesting that the photosphere is maximally aspherical at this time, likely arising from an underlying aspherical CSM component from which narrow H and He emission are observed -- which we attribute to the stellar envelope. 

The absence of early-time X-ray detections (see Section~\ref{sec:observations}) likely suggests that any X-ray emission from ejecta-CSM interaction is unable to escape at early times, due to the opacity (density) of the CSM surrounding SN\,2024cld. Although no early-time X-ray data is available for SN\,1998S, \citet{Ofek2013_shocks} show a marginal (2$\sigma$) detection for PTF11iqb at +24\,d, inferring a luminosity of $\approx7.9\times10^{40}$ erg/s. We compare the non-detections of SN\,2024cld to this in Figure~\ref{fig:SN2024cld_Swift}, alongside some well-observed X-ray detected core collapse SNe, SN\,2023ixf~\citep{Chandra2015}, and SN\,2010jl~\citep{Nayana2025}. Our early-time observations for SN\,2024cld show no clear detection at the level of PTF11iqb under similar assumptions about the photon energy distribution. In conjunction with our peak bolometric luminosity, the non-detection places upper bounds on the ratio of X-ray luminosity to bolometric luminosity, with a conservative upper limit of $L_X / L_{bol} \lesssim 0.005$ at +10\,d post-explosion, and $L_X / L_{bol} \lesssim 0.07$ at +220\,d post-explosion. At early time such a non-detection (and strong constraint on $L_X / L_{bol}$)  is unsurprising, as the CSM is optically thick to X-rays, even if we know they are being necessarily generated in the ejecta-CSM shock interface~\citep[e.g.][]{Ofek2013_shocks}. That we do not see emission at our final \textit{Swift} epoch is more interesting -- whether this is due to an intrinsically low X-ray luminosity, or again a consequence of the optical depth of the surrounding CSM is hard to disentangle given no actual detections. A more detailed treatment with CSM modelling is beyond the scope of this paper, but we note here that further deep X-ray observations of 98S-like SNe will prove diagnostic in understanding the late-time interaction.

Post-peak, we see an approximately linear decline in the light curve of SN\,2024cld across all bands, declining as expected more quickly in the bluer bands, in common with SN\,1998S (see Figure~\ref{fig:lightcurve_comparison}). SN\,2024cld then enters a marked plateau phase, showing roughly constant photospheric temperatures ($\approx5000\,$K) and a stall in the expansion of the photospheric radius -- at least partially driven by the traditional IIP-style plateau powered by hydrogen recombination~\citep[e.g.][]{Grassberg1971,Arnett1980}. There is likely significant contribution from interaction, owing to narrow-line emission, and the declining temperature during this phase (see Figure~\ref{fig:bolometric_lc}). The early-time ($\lesssim14$\,d) light curve shows a clear excess over a typical SN\,IIP light curve, driven by extra luminosity from interaction with the inner dense CSM component. Beyond this phase there is still evidence for CSM interaction via the H$\alpha$ emission, thus it is not straightforward to infer any explosion properties based on the plateau luminosity.

As alluded to in Section~\ref{sec:spectroscopy}, additional red and blue components in the H$\alpha$ profile emerge beyond +80\,d, at velocities of order \qty{6000}{\kilo\meter\per\second}. The source of these is ejecta interacting with an aspherical CSM, with the asymmetry between the red and blue components likely arising from a non-axisymmetric mass distribution in the CSM. The most commonly invoked geometry for this is a CSM disk/torus. CSM disks have been invoked in a number of SNe previously \citep[e.g.][]{Hoffman2008,Pursiainen2023_bsz,Reynolds2025}, including in SN\,1998S~\citep{Leonard2000} and PTF11iqb~\citep{Smith2015}, with recent theoretical work~\citep{Scherbak2025} providing plausible mechanisms to generate such disks via binary interaction. The high velocity of these components is well-explained by CSM interacting with the fast ejecta, accelerating it, but the fact that we see both red and blue components suggests our viewing angle is somewhat off-centre. The early-time polarisation is inconsistent with a disk, but the rising polarisation and rotation of polarisation angle suggest the photosphere transitions between two distinct CSM components -- an inner dense envelope, and a more elongated CSM disk -- around the conclusion of the FI phase. There is also evolution of the narrow-line P Cygni feature, which further feeds into the complex velocity components present.

As the photosphere recedes back through the CSM (see Figure~\ref{fig:bolometric_lc}), we see further evolution in the relative strengths of the profiles -- though this is naturally superimposed on a growing broad H$\alpha$ component. The blue component begins to decay prior to the red, though the red component too begins to decay at late time. No real evidence for dust at this epoch from the NOTCam $JHK$ imaging is seen (see Figure~\ref{fig:bb_fits}), and thus this is likely intrinsic to the CSM. The same bulk behaviour surrounding the late-time spectral evolution is seen in SN\,1998S and PTF11iqb (see Figure~\ref{fig:spectral_comparison}), with emerging asymmetric H$\alpha$ components beyond the first few months. There are differences between the strengths of forbidden emission lines between all three 98S-likes we consider here: in SN\,198S, the [Ca II] line is clearly visible from around \qty{+100}{\day}, where in SN\,2024cld only a broad flat-topped feature is visible from \qty{+145}{\day} onwards, and no obvious [Ca II] emission in PTF\,11iqb at any epoch. Assuming the underlying progenitors are similar, this is likely indicative of the relative CSM densities between the objects, though a detailed treatment of this is beyond the scope of the paper, requiring tailored modelling.

The velocity offsets of the red and blue H$\alpha$ components are broadly comparable, with detailed differences in the H$\alpha$ profile being driven by specific differences in the spatial density/clumping of the CSM.

The drop around \qty{\approx100}{\day} signals the end of the H recombination and interaction-powered phase, likely driven by the recombination/cooling front ceasing to move back through the ejecta, as in SNe~IIP. For the remainder of our photometry, SN\,2024cld shows an extended late-time excess in luminosity, beginning around \qty{110}{\day}, and only declining by 1\,mag over this phase, reminiscent of a secondary \say{plateau}. This is in stark contrast to the typical 2-3 mag seen in SNe~IIP. This extends through the phase where SN\,2024cld becomes Sun-constrained. Given the poor availability of photometry around this phase for the other 98S-like SNe, it is unclear whether the duration of this late-time excess is comparable, but it does at least seem that the $R$-band absolute magnitude during this phase is similar between PTF11iqb and SN\,2024cld, suggesting similar CSM density at this phase. SN\,1998S appears to show a similar late-time excess in luminosity but available photometry runs out around +120d. The end of this plateau phase is poorly constrained in PTF11iqb.
The spectra obtained at these phases are clearly dominated by H$\alpha$ emission (and to a lesser extent the Ca~II NIR triplet), thus the overall luminosity at late time is likely fuelled by ongoing CSM interaction. We see additional components to H$\alpha$ at high velocities, with clear asymmetry in the strengths of the red and blue components. That we still see the photospheric features and the light curve does not drop quickly further confirms this: there must be some additional energy keeping the ejecta hot, thus visible -- plausibly heating from ongoing ejecta-CSM interaction.

\subsection{Mass loss rates and history}
Deriving precise mass loss rates for the progenitor of SN\,2024cld is challenging, given that the different phases through which the light curve evolved were likely  powered by varying combinations of CSM-ejecta interaction, shock ionisation, H recombination, and underlying radioactive decay. Nevertheless, we can pick out a few key phases where we can be reasonably confident about the CSM properties to make headway. The late-time light curve is dominated by interaction, declining markedly more slowly than the expected decline rate for $^{56}$Co decay \citep[with complete gamma-ray trapping, see ][]{Woosley1989}, as shown in Figure~\ref{fig:lightcurve_comparison}, and thus provides a route to estimate the mass loss rate required to produce the CSM driving this. Following \citet{Smith2017}, we assume that the bolometric luminosity at this epoch is driven by shock interaction with a dense CSM, and proceed to compute the wind density parameter and mass loss rates under this assumption, based on observed features in the spectra. The luminosity follows the form
$$L = \frac{1}{2}\dot{M}_\mathrm{CSM}\left(\frac{v_\mathrm{CDS}^{3}}{v_\mathrm{CSM}}\right)$$
in the case of the ejecta traversing a spherical CSM shed from e.g. the progenitor wind. Although we know that SN\,2024cld has a more complex geometry than this, the mass loss rates derived via this approach are still illustrative, even if not precise. The velocity of the cold dense shell, $v_{\mathrm{CDS}}$, is estimated from the velocities of the red and blue components of H$\alpha$ at later times (+100d and onwards, \qty{\approx 5500}{\kilo\meter\per\second}), and the CSM velocity $v_{\mathrm{CSM}}$ from the narrow H$\alpha$ P Cygni profile velocities (see Figure~\ref{fig:absorption_vel}). Mass loss rates over this late-time phase are $\approx6\times10^{-4}$ $\mathrm{M_\odot}\,\text{y}^{-1}$ (comparable to the $10^{-4}$ $\mathrm{M_\odot}\,\text{y}^{-1}$ inferred for PTF11iqb in \citealt{Smith2015}), with uncertainties of 0.3 dex. If $v_{\mathrm{CDS}}$ is overestimated (e.g. these components trace the ejecta velocity), the inferred mass loss rate could be markedly higher.

The time prior to explosion that this CSM was shed can be estimated as $\delta t \frac{v_{\mathrm{CDS}}}{v_{\mathrm{CSM}}}$, here spanning 25-50 years pre-explosion. Integrating the inferred mass loss rates from +100d to the end of our spectral series suggests 0.02 $\mathrm{M_\odot}$ was lost from the progenitor of SN\,2024cld during this time. There are no direct observations of the CDS velocity at early times as it has not yet formed, and given both the strong scaling of $\dot{M}$ with $v_{\mathrm{CDS}}$ and the rise not being fully interaction-powered, we do not apply the same formalism at these earlier phases. Nevertheless, the comparisons to flash-ionisation models (see Figure~\ref{fig:flash-features-compare}) point towards enhanced mass loss rates in the progenitor close to the time of explosion, around 5-10$\times$ higher than inferred for the historical mass loss of the progenitor probed by late-time observations of the SN. This lends further credence to the dense CSM inferred from both the inflated photosphere size at time of explosion, and the high velocity CSM inferred from the narrow H$\alpha$ P Cygni profile. However, these models are derived under the assumption of wind-driven mass loss, which we disfavour over an impulsive ejection/binary mass transfer event proximate to explosion -- making the mass loss rates inferred from these models only approximate, given the CSM likely had higher velocity, and we know from spectroscopy that the CSM was aspherical.

The evolution of the narrow H$\alpha$ at later times is well-explained by the SN ejecta sweeping through this wind-like component from the progenitor -- with the ever-decreasing wind velocity driven by the closer-in, faster wind being swept up. The final wind velocity observed (\qty{\approx60}{\kilo\meter\per\second}) -- is higher than expected for RSG winds. The continuing evolution to lower velocities implies the wind has potentially not yet reached its terminal velocity. More specifically, the ejecta has not yet swept up the faster wind closer-in to the progenitor. This implies that a H-rich CSM is still present in sufficient density beyond +200d post-explosion to drive the absorption components observed, even if some of the emission is attributable to underlying host lines. We see no evidence for any absorption in the CALIFA explosion site spectrum, taken pre-explosion.

\subsection{Progenitor system}
Obtaining a robust identification for the progenitor of SN\,2024cld is challenging -- there is no pre-explosion imaging that would provide sufficiently constraining upper limits, and at the time of our last spectrum, SN\,2024cld remains photospheric, thus making modelling of nebular features impossible until later. Nevertheless, we can consider which progenitor systems are likely capable of generating the observed CSM configuration, and proceed accordingly.

Observations of the most massive stars in the Milky Way and its satellites have revealed complex CSM configurations~\citep{Schuster2006} such as shells, rings, and asymmetric clumps. Recent millimetre-wave observations of NML Cyg~\citep{DeBeck2025} reveal a series of low (\qty{\approx30}{\kilo\meter\per\second}) velocity shells surrounding the star, presumably ejected in a sequence of eruptions a few hundred years apart. A number of evolved RSGs show resolved ring nebulae, with a prominent example of this being the \object{Fried egg nebula} surrounding \object{Hen\,3-1379}~\citep{Hutsemekers2013,Koumpia2020}. The emission geometry of the ring presents a highly aspherical photosphere, with expansion velocities typically around \qty{\approx60}{\kilo\meter\per\second}~\citep{Lagadec2011}. Both NML Cyg and Hen 3-1379 are not (currently) known to have binary companions, thus single star evolution may provide a route to generate this dense, structured CSM -- though the precise mechanisms for such mass loss in RSGs and evolved RSGs remain debated~\citep{Beasor2020}. If indeed arising from single stars, the complex CSM observed in SN\,2024cld provides further constraints on the timescale and geometry of such mass loss mechanisms. There is also the complication that the CSM in many of these systems is dominated by cool dust, rather than gas -- this would likely manifest in as strong NIR excess if present in reasonable amounts, which we do not (at least thus far) observe in the case of SN\,2024cld even at \qty{220}{\day} post-explosion -- yet was seen in both SN\,1998S~\citep{Fassia2000} and SN\,2013fc~\citep{Kangas2016}. Pre-existing dust close-in to the progenitor would likely be destroyed regardless through ejecta-CSM interaction. In both the case of \object{Hen 3-1379} and NML Cyg, it is not clear how constraining existing observations are regarding the presence of an unseen, potentially low-mass companion -- but given the star is massive, it is not unlikely this system is also binary in nature.

Another plausible progenitor scenario is from a RSG or yellow supergiant (YSG) in a binary system. Given that massive stars are overwhelmingly expected to be in binary/multiple star systems~\citep{Sana2012}, and that SN\,2024cld is clearly a CCSN, it is \textit{a priori} likely that the progenitor star is joined by at least one companion. Binary systems provide a number of additional pathways to generate the required mass loss -- through binary shedding events, and common envelope evolution phases. Whilst it is clear this pathway introduces considerably more degrees of freedom, formation of asymmetric CSM is easier in such systems. \citet{Scherbak2025} demonstrate through simulations that such binary mass loss can occur rapidly, and yield CSM of sufficient density to drive strong interaction. 

The star \object{WOH\,G64}~\citep{Westerlund1981}, in the Large Magellanic Cloud, presents one potential realisation of such a scenario, with a number of pointed similarities to the putative progenitor of SN\,2024cld. WOH\,G64 has been the subject of intensive study, as among the most extreme RSGs~\citep[e.g. ][]{Levesque2009} in the LMC, in terms of luminosity and size. Recent VLTI observations~\citep[see Fig. 1 of][]{Ohnaka2024} in $K$-band reveal a complex multi-component CSM structure: a central compact spheroidal emission, surrounded by a broad ring of fainter emission roughly $30\,R_\star$ along its major axis, and $20\,R_\star$ along its minor axis, projected on the sky plane, although the ring is close to face-on. \citet{Ohnaka2024} also report hints of an even more distant ring component, but cannot rule out that this is an artefact of the reconstruction. These components are attributed to warm CSM. The inner spheroidal CSM component of WOH G64 has a ratio of axis lengths $E\approx0.7$, which corresponds to a maximum polarisation of $p=2\%$ assuming a scattering dominated photosphere and an oblate spheroidal geometry~\citep{Hoflich1991}. This is comparable to the polarisation degree seen in SN\,2024cld at its maximum, around the time of the narrow flash lines disappearing. The inner dense material, and outer ring appear to have different (projected) orientations in the reconstructed image, which resembles the rotation in polarisation angle that was observed in SN\,2024cld between the early and latter phases. Spectrally, strong H and Ca emission is seen in the most recent spectrum of WOH G64, with a marked P Cygni profile indicating outflows are present. The velocities of these outflows are \qty{\approx80}{\kilo\meter\per\second}~\citep{Munoz-Sanchez2024}, broadly comparable with the outflow velocities seen in Figure~\ref{fig:absorption_vel}. 

WOH G64 is suggested as a symbiotic B[e] binary~\citep{Munoz-Sanchez2024} -- given the peculiar emission lines, and the observed radial velocity shifts between the various spectral epochs -- with the primary best matching a yellow hypergiant (YHG), where a significant amount of the outer envelope was lost during the transition from RSG to YHG. Whether ejected via some dynamical process in the primary~\citep[e.g.][]{Fuller2017,Morozova2020}, or stripped in one or more binary shedding events, there are strong parallels between the CSM inferred for SN\,2024cld and the CSM observed in WOH G64. Continued time-resolved observations of \object{WOH G64} and similar systems would provide estimates of the expansion velocity of the ring/disk-like structure, which would be then directly comparable to velocities inferred for the aspherical CSM seen in 98S-like SNe -- more concretely connecting the progenitors of such objects to those seen in the local Universe.  

\begin{figure}
    \centering
    \includegraphics[width=1.0\linewidth]{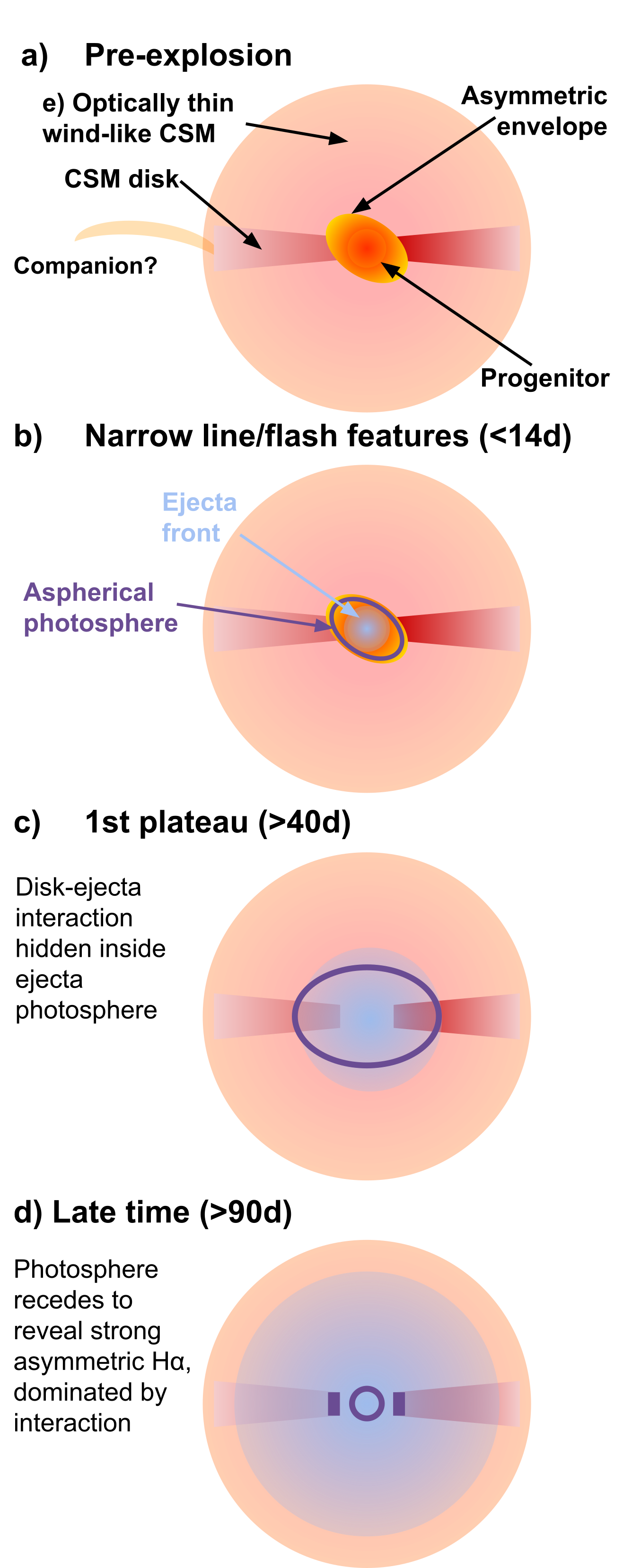}
    \caption{Summary of the inferred CSM structure around the progenitor of SN\,2024cld, alongside key observed phases in the evolution of the SN itself. Each sub-panel is discussed in the text.}
    \label{fig:scenario}
\end{figure}

We depict the key elements of the scenario we propose for SN\,2024cld, alongside their observable evidence in Figure~\ref{fig:scenario}.
\begin{enumerate}[label=\textbf{\alph*)}]
    \item SN\,2024cld likely has an evolved RSG/YHG progenitor, that has recently undergone a period of envelope inflation or impulsive mass loss, yielding an asymmetric, puffed-up geometry. The progenitor has a significant dense disk-like CSM component, with some azimuthal asymmetries -- likely caused by interaction with a binary companion, or rotation~\citep{Maeder2002}. The entire system is embedded in a tenuous wind-like environment, shed from the progenitor before the disk/envelope inflation phase.
    \item At early times, the ejecta front has not yet fully swept the inner photo-ionised envelope, and the photosphere lies ahead of the ejecta front, driving narrow line emission at early phases. Given the asymmetry, the photosphere itself shows some emerging ellipticity around \qty{7}{\day} post-explosion.
    \item During the early plateau phase, the CSM disk component is fully concealed within the photosphere associated with the ejecta. We see a largely symmetric H$\alpha$ profile during this phase, with ejecta features in absorption.
    \item At late times, the photosphere recedes in mass coordinate (albeit not significantly) back through the ejecta, revealing the strong asymmetric H$\alpha$ profiles originating from the CSM disk. Given the fact we see no strong nebular emission and both ejecta and CSM features, the photosphere cannot have receded as far as in typical CCSNe.
    \item Throughout the evolution of SN\,2024cld, we see a narrow P Cygni feature, due to the wind-like CSM throughout the system. The velocity of the absorption decreases over time as the ejecta sweeps up the faster wind closer in to the progenitor.
\end{enumerate}
The picture we paint here of the SN\,2024cld progenitor system agrees well with that of PTF11iqb presented in \citet{Smith2015} -- an extended progenitor star as inferred from the light curve rise time, most likely a RSG, embedded in a dense yet slow wind, with a prominent disk-like CSM component. The constraints on the progenitor of SN\,1998S are largely qualitative, with \citet{Fassia2001} inferring a core mass of $4\,\mathrm{M_\odot}$ and an approximate zero age main sequence (ZAMS) mass of $20\,\mathrm{M_\odot}$ -- a massive progenitor. A similar mass was also inferred for the progenitor of SN\,2013fc~\citep{Kangas2016}. The late-time H$\alpha$ luminosity \citep{Mauerhan2012} is also consistent with a RSG based on the models of \citet{Chevalier1994}.
\citet{Leonard2000} suggest, based on their spectropolarimetry, that SN\,1998S showed (at least) two distinct mass loss episodes -- one with a ring/disk-like morphology that concluded in the final few years of the progenitor's life, and another that was still occurring as the star exploded, which yielded a confined CSM close-in to the progenitor. This matches well the CSM geometry we infer for SN\,2024cld, with differences in polarisation degree and evolution attributable to viewing angle effects. The class of 98S-like events (including SN\,2024cld) all seem to have remarkably similar CSM geometries which would, at least at face value, imply very similar mass loss pathways between them, and thus progenitor systems. Given the intrinsic rarity of the 98S-like SNe, this implies that the progenitor channel that forms them cannot be very common. The presence of disk-like CSM is of course not unique to SN\,2024cld and the 98S-likes, being invoked in a number of recent interacting H-rich SNe~\citep[e.g. ][]{Andrews2025,Tartaglia2025,Charalampopoulos2025,Nagao2025,Reynolds2025}, -- yet SN\,2024cld probes a different level of interaction/pre-explosion mass loss to these examples, and the dataset presented here is among the most comprehensive examples to date.

Regardless of how the CSM was shed, the observed properties of SN\,2024cld point towards a moderately massive H-rich progenitor star that underwent a period of extreme mass transfer (whether via single or binary mechanism) in the final few hundred years prior to explosion, and became greatly inflated in the final few years of its life. This is in line with suggestions for the similar 98S-like objects, with evolved RSG progenitors undergoing significant mass loss. The qualitative differences between the light curve and spectra of each object are entirely consistent with differing mass loss profiles over time -- with SN\,1998S showing a more IIL-like light curve owing to a greater degree of mass loss at late times throwing off the H-rich material needed for a recombination-fuelled plateau, and PTF~11iqb showing a more IIP-like light curve. The  late-time light curve of SN\,2024cld does not directly match either object, but is more similar to PTF11iqb overall. The continued late-time interaction-driven lightcurve and pre-shock CSM features of SN\,2024cld suggests that the CSM is highly extended out to scales of $\sim10^{16}$\,cm. Continued spectroscopic observations at late time will further probe the geometry of mass loss, including components shed earlier in the progenitor evolution that the ejecta has not yet reached.

\section{Conclusions} \label{sec:conclusions}
In this paper, we presented the UV/optical/IR follow-up campaign on SN\,2024cld, a transitional SN~II with strong similarities to PTF~11iqb and SN\,1998S, discovered and classified just 12h post-explosion via the high-cadence GOTO-FAST survey. The densely-sampled spectral series for the object reveals persistent interaction signatures across multiple distinct phases: early-time flash-ionisation dominated emission, through to late-time asymmetric line profiles arising from ejecta interacting with an aspherical CSM, and a persistent narrow H$\alpha$ component associated with pre-shock CSM. The light curve further underscores this, showing a moderately extended rise, an early plateau, and strong late-time flattening in the light curve, the latter likely powered by interaction. A comprehensive polarimetric series charts the evolving geometry of the photosphere up to 115 days post-explosion -- showing an initially spherical configuration, that reaches a peak polarisation of $\approx2\%$ and exhibits a rotation of $\approx60^{\circ}$, before declining back towards low polarisation at late time.
Taken together, SN\,2024cld exhibits evidence for multiple CSM components: a dense, asymmetric envelope of H-rich material close-in to the progenitor, with a disk/torus-like CSM component extending out to large radii, and a strong wind-like component visible along the line of sight. The complexity of mass loss required to generate these observables further points towards a binary origin for SN\,2024cld: with discoidal/toroidal CSM in particular best explained through a binary shedding event, or potentially multiple distinct events. The CSM structuring shows marked similarity to that observed around the Galactic RSG \object{WOH~G64}~\citep{Ohnaka2024}, itself thought to have a binary companion. Assuming its progenitor is also a similar (evolved) RSG, SN\,2024cld likely represents one realisation of the explosive endpoint of such systems. Obtaining pre-explosion imaging of the progenitor of a 98S-like SN would serve to make concrete this association.

Continued spectroscopic observations of SN\,2024cld out to late times remain crucial: to constrain the edge of the asymmetric CSM and thus the time it was shed, continue to monitor the pre-shock CSM wind velocity and thus measure the progenitor wind, and to probe the progenitor mass via the nebular emission, when (if) it emerges before the transient becomes too faint.

\section*{Acknowledgements}
We thank the referee, Jennifer Andrews, for her insightful comments and suggestions.
TLK acknowledges support via an Research Council of Finland grant (340613), support from the Turku University Foundation (grant no. 081810), and a Warwick Astrophysics prize post-doctoral fellowship made possible thanks to a generous philanthropic donation.
MP, JL, DON acknowledge support from a UK Research and Innovation Future Leaders Fellowship (grant references MR/T020784/1 and UKRI1062).  
RK and PC acknowledge support via Research Council of Finland (grant 340613).
DLC acknowledges support from the Science and Technology Facilities Council (STFC) grant number ST/X001121/1.
MRM acknowledges support via a Warwick Astrophysics prize post-doctoral fellowships made possible thanks to a generous philanthropic donation.
MGB acknowledges financial support from the Spanish Ministerio de Ciencia e Innovación (MCIN) and the Agencia Estatal de Investigación (AEI) 10.13039/501100011033 under the PID2023-151307NB-I00 SNNEXT project, from Centro Superior de Investigaciones Científicas (CSIC) under the PIE project 20215AT016 and the program Unidad de Excelencia María de Maeztu CEX2020-001058-M, and from the Departament de Recerca i Universitats de la Generalitat de Catalunya through the 2021-SGR-01270 grant.
YZC is supported by the National Natural Science Foundation of China (NSFC, Grant No. 12303054), the National Key Research and Development Program of China (Grant No. 2024YFA1611603), the Yunnan Fundamental Research Projects (Grant Nos. 202401AU070063, 202501AS070078), and the International Centre of Supernovae, Yunnan Key Laboratory (No. 202302AN360001).
NER. acknowledges support from the Spanish Ministerio de Ciencia e Innovaci\'on (MCIN) and the Agencia Estatal de Investigaci\'on (AEI) 10.13039/501100011033 under the program Unidad de Excelencia Mar\'ia de Maeztu CEX2020-001058-M.
LG acknowledges financial support from AGAUR, CSIC, MCIN and AEI 10.13039/501100011033 under projects PID2023-151307NB-I00, PIE 20215AT016, CEX2020-001058-M, ILINK23001, COOPB2304, and 2021-SGR-01270.
CPG acknowledges financial support from the Secretary of Universities and Research (Government of Catalonia) and by the Horizon 2020 Research and Innovation Programme of the European Union under the Marie Sk\l{}odowska-Curie and the Beatriu de Pin\'os 2021 BP 00168 programme, from the Spanish Ministerio de Ciencia e Innovaci\'on (MCIN) and the Agencia Estatal de Investigaci\'on (AEI). 10.13039/501100011033 under the PID2023-151307NB-I00 SNNEXT project, from Centro Superior de Investigaciones Cient\'ificas (CSIC) under the PIE project 20215AT016 and the program Unidad de Excelencia Mar\'ia de Maeztu CEX2020-001058-M, and from the Departament de Recerca i Universitats de la Generalitat de Catalunya through the 2021-SGR-01270 grant.
AK is supported by the UK Science and Technology Facilities Council (STFC) Consolidated grant ST/V000853/1.
LK acknowledges support for an Early Career Fellowship from the Leverhulme Trust through grant ECF-2024-054 and the Isaac Newton Trust through grant 24.08(w).
TK acknowledges support from the Research Council of Finland project 360274.
GL was supported by a research grant (VIL60862) from VILLUM FONDEN.
SM acknowledges support from the Research Council of Finland project 350458.
AR acknowledges financial support from the GRAWITA Large Program Grant (PI P. D’Avanzo) and from the PRIN-INAF 2022 \say{Shedding light on the nature of gap transients: from the observations to the models}.
TMR is part of the Cosmic Dawn Center (DAWN), which is funded by the Danish National Research Foundation under grant DNRF140. TMR acknowledges support from the Research Council of Finland project 350458. 
AS acknowledges a Warwick Astrophysics PhD prize scholarship made possible thanks to a generous philanthropic donation.
IS acknowledges financial support from the Son-Of-X-Shooter Consortium and the PRIN-INAF 2022 \say{Shedding light on the nature of gap transients: from the observations to the models}.
DS acknowledges support from the Science and Technology Facilities Council (STFC), grant numbers ST/T007184/1, ST/T003103/1, ST/T000406/1, ST/X001121/1 and ST/Z000165/1.

The Gravitational-wave Optical Transient Observer (GOTO) project acknowledges the support of the Monash-Warwick Alliance; University of Warwick; Monash University; University of Sheffield; University of Leicester; Armagh Observatory \& Planetarium; the National Astronomical Research Institute of Thailand (NARIT); Instituto de Astrofísica de Canarias (IAC); University of Portsmouth; University of Turku; University of Birmingham; and the UK Science and Technology Facilities Council (STFC, grant numbers ST/T007184/1, ST/T003103/1 and ST/Z000165/1).
The Isaac Newton Telescope is operated on the island of La Palma by the Isaac Newton Group of Telescopes in the Spanish Observatorio del Roque de los Muchachos of the Instituto de Astrofísica de Canarias, with observations gathered under proposal I/2024A/07.
The NOT data presented here were obtained with ALFOSC, which is provided by the Instituto de Astrofisica de Andalucia (IAA) under a joint agreement with the University of Copenhagen and NOT.
We acknowledge funding to support our NOT observations from the Finnish Centre for Astronomy with ESO (FINCA), University of Turku, Finland (Academy of Finland grant nr 306531).
Based on observations collected at Copernico 1.82m telescope and Schmidt 67/92 telescopes (Asiago Mount Ekar, Italy) INAF - Osservatorio Astronomico di Padova.
This work has made use of data from the Asteroid Terrestrial-impact Last Alert System (ATLAS) project. The Asteroid Terrestrial-impact Last Alert System (ATLAS) project is primarily funded to search for near earth asteroids through NASA grants NN12AR55G, 80NSSC18K0284, and 80NSSC18K1575; byproducts of the NEO search include images and catalogs from the survey area. This work was partially funded by Kepler/K2 grant J1944/80NSSC19K0112 and HST GO-15889, and STFC grants ST/T000198/1 and ST/S006109/1. The ATLAS science products have been made possible through the contributions of the University of Hawaii Institute for Astronomy, the Queen’s University Belfast, the Space Telescope Science Institute, the South African Astronomical Observatory, and The Millennium Institute of Astrophysics (MAS), Chile.
Based on observations made with the Nordic Optical Telescope, owned in collaboration by the University of Turku and Aarhus University, and operated jointly by Aarhus University, the University of Turku and the University of Oslo, representing Denmark, Finland and Norway, the University of Iceland and Stockholm University at the Observatorio del Roque de los Muchachos, La Palma, Spain, of the Instituto de Astrofisica de Canarias.
Based in part on data acquired at the ANU 2.3-metre telescope (under program 2410086). The automation of the telescope was made possible through an initial grant provided by the Centre of Gravitational Astrophysics and the Research School of Astronomy and Astrophysics at the Australian National University and through a grant provided by the Australian Research Council through LE230100063. The Lens proposal system is maintained by the AAO Research Data \& Software team as part of the Data Central Science Platform. We acknowledge the traditional custodians of the land on which the telescope stands, the Gamilaraay people, and pay our respects to elders past and present.
We acknowledge the use of public data from the Swift data archive.
This study makes uses of the data provided by the Calar Alto Legacy Integral Field Area (CALIFA) survey (\url{https://califa.caha.es/}), and is partially based on observations collected at the Centro Astronómico Hispano Alemán (CAHA) at Calar Alto, operated jointly by the Max-Planck-Institut fűr Astronomie and the Instituto de Astrofisica de Andalucia (CSIC).
Funding for SDSS-III has been provided by the Alfred P. Sloan Foundation, the Participating Institutions, the National Science Foundation, and the U.S. Department of Energy Office of Science. The SDSS-III web site is http://www.sdss3.org/.
SDSS-III is managed by the Astrophysical Research Consortium for the Participating Institutions of the SDSS-III Collaboration including the University of Arizona, the Brazilian Participation Group, Brookhaven National Laboratory, Carnegie Mellon University, University of Florida, the French Participation Group, the German Participation Group, Harvard University, the Instituto de Astrofisica de Canarias, the Michigan State/Notre Dame/JINA Participation Group, Johns Hopkins University, Lawrence Berkeley National Laboratory, Max Planck Institute for Astrophysics, Max Planck Institute for Extraterrestrial Physics, New Mexico State University, New York University, Ohio State University, Pennsylvania State University, University of Portsmouth, Princeton University, the Spanish Participation Group, University of Tokyo, University of Utah, Vanderbilt University, University of Virginia, University of Washington, and Yale University.
This work has made use of data from the European Space Agency (ESA) mission
{\it Gaia} (\url{https://www.cosmos.esa.int/gaia}), processed by the {\it Gaia}
Data Processing and Analysis Consortium (DPAC,
\url{https://www.cosmos.esa.int/web/gaia/dpac/consortium}). Funding for the DPAC has been provided by national institutions, in particular the institutions participating in the {\it Gaia} Multilateral Agreement.
We acknowledge ESA Gaia, DPAC and the Photometric Science Alerts Team (\url{http://gsaweb.ast.cam.ac.uk/alerts}).

\section*{Data Availability}
A machine-readable version of Table~\ref{tab:photlog} will be given as part of the Supplementary Materials upon acceptance.
All spectroscopy presented in this work will be made available via WISeREP (\url{https://wiserep.org}) upon acceptance.
All other data is available from the corresponding Author upon reasonable request.

\bibliographystyle{mnras}
\bibliography{refs}
\smallskip

\hrule
\smallskip
\noindent
{
\footnotesize
$^{1}$Department of Physics, University of Warwick, Gibbet Hill Road, Coventry CV4 7AL, UK.\\
$^{2}$Department of Physics and Astronomy, University of Turku, 20014 Turku, Finland.\\
$^{3}$Finnish Centre for Astronomy with ESO (FINCA), University of Turku, 20014 Turku, Finland.\\
$^{4}$School of Physics \& Astronomy, Monash University, Clayton VIC 3800, Australia.\\
$^{5}$Astrophysics Research Cluster, School of Mathematical and Physical Sciences, University of Sheffield, Sheffield S3 7RH, UK.\\
$^{6}$Research Software Engineering, University of Sheffield, Sheffield, S1 4DP, UK.\\
$^{7}$Institute of Space Sciences (ICE, CSIC), Campus UAB, Carrer de Can Magrans s/n, 08193 Barcelona, Spain.\\
$^{8}$Institut d’Estudis Espacials de Catalunya (IEEC), 08860 Castelldefels (Barcelona), Spain.\\
$^{9}$Department of Physics, Royal Holloway – University of London, Egham Hill, Egham TW20 0EX, UK.\\
$^{10}$School of Physics and Astronomy, University of Birmingham, Edgbaston, Birmingham, B15 2TT, UK.\\
$^{11}$Institute for Gravitational Wave Astronomy, University of Birmingham, Birmingham, B15 2TT, UK.\\
$^{12}$Jodrell Bank Centre for Astrophysics, Department of Physics and Astronomy, The University of Manchester, Manchester M13 9PL, UK.\\
$^{13}$Centre for Advanced Instrumentation, University of Durham, DH1 3LE Durham, UK.\\
$^{14}$Yunnan Observatories, Chinese Academy of Sciences, Kunming, 650216, P.R. China.\\
$^{15}$International Centre of Supernovae, Yunnan Key Laboratory, Kunming 650216, P.R. China.\\
$^{16}$Instituto de Astrofísica de Canarias, E-38205 La Laguna, Tenerife, Spain.\\
$^{17}$INAF-Osservatorio Astronomico di Padova, Vicolo dell’Osservatorio 5, 35122 Padova, Italy.\\
$^{18}$School of Physics, University College Dublin, L.M.I. Main Building, Beech Hill Road, Dublin 4 D04 P7W1, Ireland.\\
$^{19}$Institute of Astronomy and Kavli Institute for Cosmology, University of Cambridge, Madingley Road, Cambridge CB3 0HA, UK.\\
$^{20}$DTU Space, Department of Space Research and Space Technology, Technical University of Denmark, Elektrovej 327, 2800 Kgs. Lyngby, Denmark.\\
$^{21}$School of Sciences, European University Cyprus, Diogenes street, Engomi, 1516 Nicosia, Cyprus.\\
$^{22}$Aalto University Metsähovi Radio Observatory, Metsähovintie 114, 02540 Kylmälä, Finland.\\
$^{23}$Aalto University Department of Electronics and Nanoengineering, PO Box 15500, 00076 Aalto, Finland.\\
$^{24}$National Astronomical Research Institute of Thailand, 260 Moo 4, T. Donkaew, A. Maerim, Chiangmai 50180, Thailand.\\
$^{25}$Institute of Cosmology and Gravitation, University of Portsmouth, Portsmouth PO1 3FX, UK.\\
$^{26}$School of Physics \& Astronomy, University of Leicester, University Road, Leicester LE1 7RH, UK.\\
$^{27}$Armagh Observatory and Planetarium, College Hill, Armagh, BT61 9DG, N Ireland, UK.\\
$^{28}$INAF – Osservatorio Astronomico di Brera, Via E. Bianchi 46, I-23807 Merate (LC), Italy.\\
$^{29}$Tuorla Observatory, Department of Physics and Astronomy, 20014, University of Turku, Vesilinnantie 5, Turku, Finland.\\
$^{30}$Cosmic DAWN Centre.\\
$^{31}$Niels Bohr Institute, University of Copenhagen, Jagtvej 128, 2200 København N, Denmark.\\
$^{32}$INAF–Osservatorio Astronomico di Capodimonte, Salita Moiariello, 16, 80131 Napoli, Italy.\\
$^{33}$Department of Physics and Astronomy, Aarhus University, Ny Munkegade 120, DK-8000 Aarhus C, Denmark.\\
$^{34}$School of Physics and Astronomy, Beijing Normal University, Beijing 100875, China.\\
$^{35}$Department of Physics, Faculty of Arts and Sciences, Beijing Normal University, Zhuhai 519087, China.\\
}
\appendix
\section{Additional tables and figures}

\FloatBarrier
\begin{figure*}
    \centering
    \includegraphics[width=0.5\linewidth]{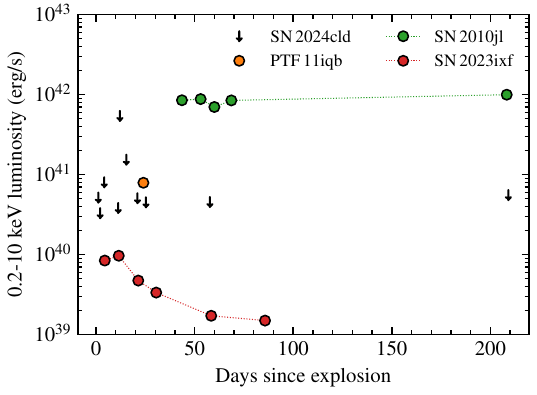}
    \caption{Luminosities in the 0.2-10 keV band for a number of X-ray detected SNe, alongside our derived $2\sigma$ non-detections for SN\,2024cld.}
    \label{fig:SN2024cld_Swift}
\end{figure*}
\begin{figure*}
    \centering
    \includegraphics[width=0.7\linewidth]{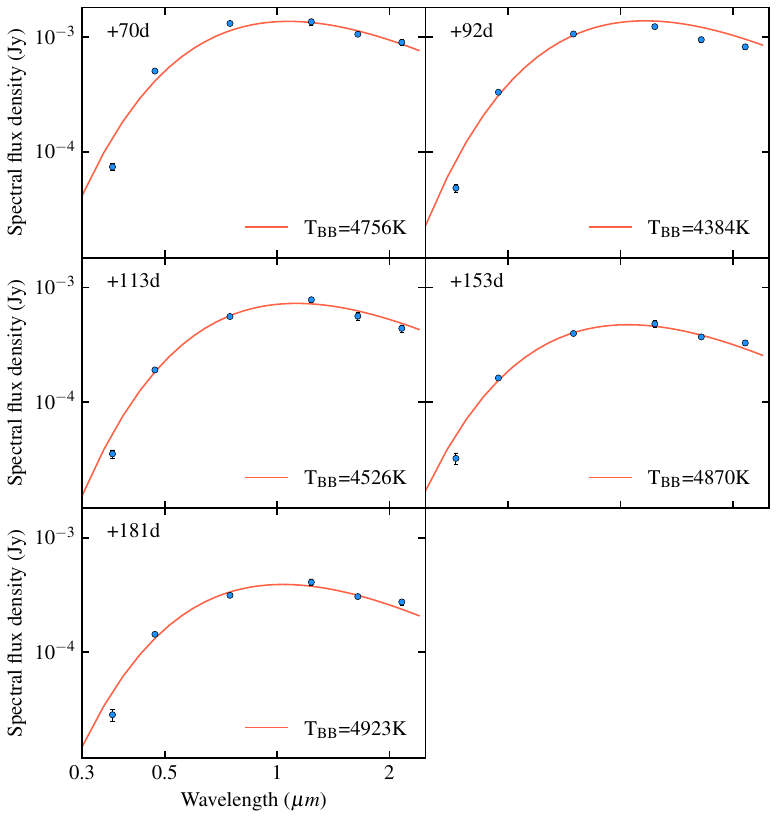}
    \caption{Blackbody fits to the observed photometry of SN\,2024cld at the epochs for which we have $JHK$-band measurements available. The $r$ and $z$ bands are excluded from the fit due to strong H$\alpha$ and Ca~II emission respectively. All epochs are fit well by a single blackbody component, suggesting no warm dust is present in the system, unlike SN\,1998S~\citep{Fassia2000} and SN\,2013fc~\citep{Kangas2016} at similar phase.}
    \label{fig:bb_fits}
\end{figure*}
\begin{figure}
    \centering
    \includegraphics[width=\linewidth]{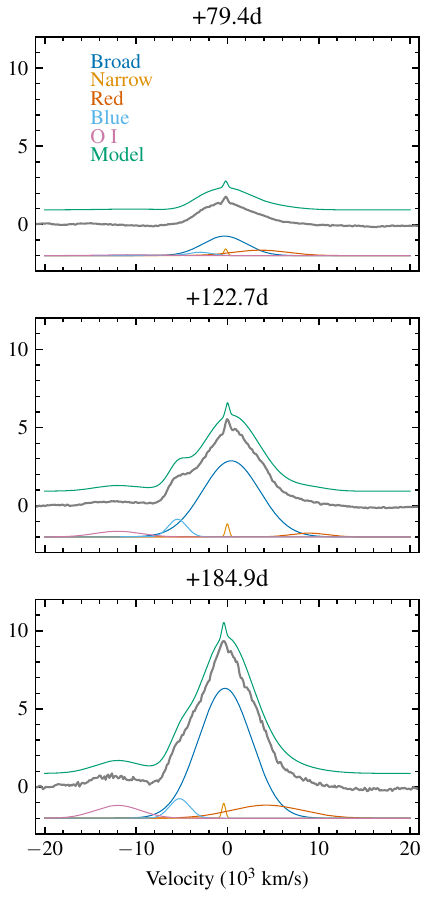}
    \caption{Results of Gaussian fitting to the complex H$\alpha$ profile observed in SN\,2024cld for three indicative epochs of observation. The composite model is overplotted on the normalised flux, alongside individual model components (offset by a constant value for clarity).}
    \label{fig:halpha_fits}
\end{figure}
\begin{table*}
\caption{Log of photometry from triggered follow-up of SN\,2024cld.The full machine-readable version of this table is given as part of the Supplementary Materials. All photometry is given in its' native the AB system, and uncorrected for Galactic and host reddening.}
\label{tab:photlog}
\begin{tabular}{lllllll}
\hline
Date (UT) & MJD & Phase & Band & Magnitude & Uncertainty & Telescope \\
\hline\hline
2024-02-13 & 60353.27 & 0.5 & z & 18.19 & 0.06 & LT \\
2024-02-13 & 60353.27 & 0.5 & i & 18.02 & 0.04 & LT \\
2024-02-13 & 60353.27 & 0.5 & r & 17.85 & 0.04 & LT \\
     &  &     & \vdots  &       &      &    \\
2024-09-11 & 60564.86 & 209.5 & g & 18.74 & 0.02 & NOT \\
2024-09-11 & 60564.87 & 209.5 & i & 17.85 & 0.02 & NOT \\
2024-09-11 & 60564.87 & 209.5 & z & 17.35 & 0.02 & NOT \\
\hline
\end{tabular}

\end{table*}
\begin{table*}
\caption{Log of spectral observations for SN\,2024cld. For spectra composed of multiple grating settings/spectrograph arms, exposure times correspond to the per-grating exposure time. No slit widths are given for the ANU2.3m/WiFeS given it is an integral field spectrograph, but each slitlet has a nominal width of 1\arcsec~\citep{Dopita2007}.}

\label{tab:spectralog}
\begin{tabular}{lrrlllrrr}
\hline
Date & MJD & Phase & Telescope & Instrument & Grism & Airmass & Exposure time & Slit width \\
\hline
UT   &     & d     &           &            &       &         & s             & arcsec \\
\hline\hline
2024-02-13 & 60353.25 & 0.5 & INT & IDS & R150V & 1.06 & 2400 & -- \\
2024-02-14 & 60354.17 & 1.4 & INT & IDS & R150V & 1.42 & 1800 & -- \\
2024-02-16 & 60356.15 & 3.4 & NOT & ALFOSC & gr8/18 & 1.64 & 900 & 1.3 \\
2024-02-16 & 60356.25 & 3.5 & LT & SPRAT & blue & 1.07 & 1500 & 1.8 \\
2024-02-20 & 60360.12 & 7.3 & NOT & ALFOSC & gr8/18 & 1.89 & 900 & 1.3 \\
2024-02-20 & 60360.19 & 7.4 & LT & SPRAT & blue & 1.24 & 1500 & 1.8 \\
2024-02-21 & 60361.17 & 8.3 & NOT & ALFOSC & gr8/18 & 1.35 & 900 & 1.0 \\
2024-02-24 & 60364.21 & 11.3 & LT & SPRAT & blue & 1.12 & 900 & 1.8 \\
2024-02-25 & 60365.20 & 12.3 & LT & SPRAT & blue & 1.13 & 900 & 1.8 \\
2024-02-28 & 60368.18 & 15.3 & INT & IDS & R400V & 1.13 & 2700 & -- \\
2024-02-29 & 60369.16 & 16.2 & NOT & ALFOSC & gr8/18 & 1.31 & 900 & 1.3 \\
2024-03-05 & 60374.16 & 21.2 & NOT & ALFOSC & gr8/18 & 1.24 & 900 & 1.0 \\
2024-03-08 & 60377.71 & 24.7 & ANU 2.3m & WiFeS & B7000/R7000 & 1.81 & 3200 & -- \\
2024-03-10 & 60379.18 & 26.1 & NOT & ALFOSC & gr17 & 1.09 & 900 & 0.9 \\
2024-03-10 & 60379.19 & 26.1 & NOT & ALFOSC & gr4 & 1.06 & 600 & 1.0 \\
2024-03-15 & 60384.25 & 31.1 & GTC & OSIRIS+ & R1000B/R & 1.02 & 500 & 1.0 \\
2024-03-16 & 60385.18 & 32.0 & NOT & ALFOSC & gr17 & 1.07 & 900 & 0.9 \\
2024-03-16 & 60385.19 & 32.0 & NOT & ALFOSC & gr4 & 1.05 & 600 & 1.0 \\
2024-03-25 & 60394.64 & 41.4 & ANU 2.3m & WiFeS & B3000/R3000 & 1.99 & 3200 & -- \\
2024-03-31 & 60400.16 & 46.8 & NOT & ALFOSC & gr4 & 1.03 & 600 & 1.3 \\
2024-04-01 & 60401.08 & 47.7 & NOT & ALFOSC & gr17 & 1.25 & 900 & 0.9 \\
2024-04-12 & 60412.12 & 58.7 & NOT & ALFOSC & gr4 & 1.04 & 600 & 1.3 \\
2024-04-14 & 60414.02 & 60.5 & NOT & ALFOSC & gr17 & 1.40 & 900 & 0.9 \\
2024-04-20 & 60420.95 & 67.4 & NOT & ALFOSC & gr4 & 2.11 & 900 & 1.3 \\
2024-04-27 & 60427.15 & 73.5 & NOT & ALFOSC & gr17 & 1.03 & 1200 & 0.9 \\
2024-05-03 & 60433.08 & 79.4 & NOT & ALFOSC & gr4 & 1.02 & 900 & 1.0 \\
2024-05-06 & 60436.08 & 82.3 & NOT & ALFOSC & gr4 & 1.02 & 900 & 1.0 \\
2024-05-06 & 60436.63 & 82.9 & ANU 2.3m & WiFeS & B7000/R7000 & 1.57 & 3600 & -- \\
2024-05-10 & 60440.20 & 86.4 & NOT & ALFOSC & gr17 & 1.27 & 1200 & 0.9 \\
2024-05-15 & 60445.02 & 91.1 & NOT & ALFOSC & gr4 & 1.06 & 900 & 1.0 \\
2024-05-20 & 60451.00 & 97.0 & NOT & ALFOSC & gr17 & 1.07 & 1200 & 0.9 \\
2024-05-28 & 60458.12 & 104.1 & NOT & ALFOSC & gr4 & 1.16 & 1200 & 1.0 \\
2024-06-07 & 60468.99 & 114.8 & ANU 2.3m & WiFeS & B7000/R7000 & 1.72 & 3600 & -- \\
2024-06-08 & 60469.04 & 114.9 & NOT & ALFOSC & gr4 & 1.03 & 1800 & 1.0 \\
2024-06-15 & 60476.95 & 122.7 & NOT & ALFOSC & gr4 & 1.04 & 1800 & 1.0 \\
2024-06-23 & 60484.96 & 130.6 & NOT & ALFOSC & gr4 & 1.02 & 1800 & 1.3 \\
2024-07-04 & 60495.42 & 140.9 & ANU 2.3m & WiFeS & B7000/R7000 & 1.67 & 3200 & -- \\
2024-07-08 & 60499.98 & 145.4 & NOT & ALFOSC & gr4 & 1.06 & 2400 & 1.0 \\
2024-07-19 & 60510.98 & 156.3 & NOT & ALFOSC & gr4 & 1.15 & 2400 & 1.0 \\
2024-08-01 & 60523.90 & 169.1 & NOT & ALFOSC & gr4 & 1.05 & 2400 & 1.0 \\
2024-08-17 & 60539.94 & 184.9 & NOT & ALFOSC & gr4 & 1.35 & 2400 & 1.3 \\
2024-08-18 & 60540.93 & 185.9 & NOT & ALFOSC & gr17 & 1.31 & 3600 & 0.5 \\
2024-08-31 & 60553.92 & 198.7 & NOT & ALFOSC & gr4 & 1.51 & 2400 & 1.0 \\
2024-09-11 & 60564.89 & 209.6 & NOT & ALFOSC & gr4 & 1.44 & 3600 & 1.0 \\
2024-09-12 & 60565.85 & 210.5 & NOT & ALFOSC & gr17 & 1.26 & 3600 & 0.5 \\
\hline
\end{tabular}
\end{table*}

\bsp	
\label{lastpage}
\end{document}